\documentclass[]{aastex631}

\usepackage{graphicx}
\usepackage{float}
\usepackage[caption=false]{subfig}
\usepackage{enumitem}

\newcommand{\degree}{^\circ}
\newcommand{\microarcsec}{$\mu$as}
\newcommand{\starry}{\texttt{starry}}
\newcommand{\hdb}{HD~189733~b}

\received{June 17, 2022}
\revised{February 23, 2023}

\shorttitle{Eclipse Mapping with Baseline}
\shortauthors{Schlawin et al.}

\begin{document}

\title{Planet Eclipse Mapping with Long-Term Baseline Drifts}

\correspondingauthor{Everett Schlawin}
\email{eas342 AT EMAIL Dot Arizona .edu}

\author[0000-0001-8291-6490]{Everett Schlawin}
\affiliation{Steward Observatory \\
933 North Cherry Avenue \\
Tucson, AZ 85721, USA}

\author{Ryan Challener}
\affiliation{Department of Astronomy, University of Michigan, 1085 S. University Ave., Ann Arbor, MI 48109, USA}

\author[0000-0003-4241-7413]{Megan Mansfield}
\affiliation{Steward Observatory \\
933 North Cherry Avenue \\
Tucson, AZ 85721, USA}

\author{Emily Rauscher}
\affiliation{Department of Astronomy, University of Michigan, 1085 S. University Ave., Ann Arbor, MI 48109, USA}

\author[0000-0002-7139-3695]{Arthur D. Adams}
\affiliation{Department of Earth and Planetary Sciences \\
University of California, Riverside \\
900 University Ave., Riverside, CA 92507, USA}

\author[0000-0002-0746-1980]{Jacob Lustig-Yaeger}
\affiliation{The Johns Hopkins University Applied Physics Laboratory, 11100 Johns Hopkins Rd, Laurel, MD, 20723, USA}



\begin{abstract}

High precision lightcurves combined with eclipse mapping techniques can reveal the horizontal and vertical structure of a planet's thermal emission and the dynamics of hot Jupiters.
Someday, they even may reveal the surface maps of rocky planets.
However, inverting lightcurves into maps requires an understanding of the planet, star and instrumental trends because they can resemble the gradual flux variations as the planet rotates (ie partial phase curves).
In this work, we simulate lightcurves with baseline trends and assess the impact on planet maps.
Baseline trends can be erroneously modeled by incorrect astrophysical planet map features, but there are clues to avoid this pitfall in both the residuals of the lightcurve during eclipse and sharp features at the terminator of the planet.
Models that use a Gaussian process or polynomial to account for a baseline trend successfully recover the input map even in the presence of systematics but with worse precision for the m=1 spherical harmonic terms.
This is also confirmed with the ThERESA eigencurve method where fewer lightcurve terms can model the planet without correlations between the components.
These conclusions help aid the decision on how to schedule observations to improve map precision.
If the m$=$1 components are critical, such as measuring the East/West hotspot shift on a hot Jupiter, better characterization of baseline trends can improve the m$=1$ terms' precision.
For latitudinal North/South information from m$\ne$1 mapping terms, it is preferable to obtain high signal-to-noise at ingress/egress with more eclipses.

\end{abstract}

\keywords{planets: atmospheres --- stars: individual ---
stars: variables: general}



\section{Introduction} \label{sec:intro}

Maps of exoplanets give powerful glimpses of the dynamics, chemistry and even land surfaces of new worlds, but the spatial resolution required (better than 30 \microarcsec\ for Proxima Cen b assuming it is earth-sized) is prohibitive for direct imaging (requiring an 8 km telescope operating at 1 \micron).
Instead, we may learn about the surfaces and maps of other worlds through indirect means like the rotational modulations and eclipses of other worlds.
The eclipse mapping technique uses the stellar limb as a spatial scan across the planet as the light from the planet is occulted by the star \citep{williams2006eclipseMapping,deWit2012eclipsemap189,rauscher2009eclipseMapping, majeau2012eclipsemap189}.
Maps at different wavelengths will also enable the identification of spectrally similar regions of the planet \citep{mansfield2020eigenspectra} and enable the construction a three dimensional (longitude, latitude and altitude) model of a planet \citep{challener2021ThERESA}.

Eclipse mapping can be used to measure the brightness gradients across a planets dayside as well as the position and extent of a hot Jupiter's hot spot without needing a full phase curve.
Even a single eclipse of \hdb\ with JWST has enough photons to determine its hot spot longitude to better than $\pm 3.5\deg$ if the map is known perfectly \citep{schlawin2018JWSTforecasts}.
In other words, if the forward map and recovered map are identical and the only free parameter in the recovered map is the longitudinal rotation, then the precision is $\pm 3.5 \deg$.
Eclipse Mapping opens up the possibility of mapping many planets and comes ``for free'' anytime there is an eclipse spectrum measurement of a bright target and large planet.

Typical eclipse mapping analysis assumes that the overall flux trends are well understood and only due to the planet flux.
In reality, stars are variable due to rotating spots \citep[e.g.][]{mcquillan2014rotationPeriodsAutoC} and there are non-astrophysical instrument trends such as visit-long slopes with Hubble Space Telescope \citep{wakeford2018wasp39} or the sinusoidal variations on 3-6 hour timescales from the reaction wheel heaters on the Kepler mission \citep{beichman2014pasp}.
Stellar super-granulation can also occur on similar timescales as the planet orbit \citep{lally2022hatp7reassessingVariability}, causing flux variations that can potentially mimic weather variability in a hot Jupiter like HAT-P-7 b \citep{armstrong2016hatp7variability}.
In this paper, we examine what happens when there are astrophysical or instrumental trends in a lightcurve and how this affects mapping results.
We use the \starry\ package to simulate a forward model, as described in Section \ref{sec:forwardModel}.
We then attempt to recover the input map, as described in Section \ref{sec:recoveredMap} using spherical harmonics.
We also use the new ThERESA package \citep{challener2021ThERESA} to recover the map with a set of eigencurves and associated eigenmaps.

\begin{figure*}
\gridline{
	\fig{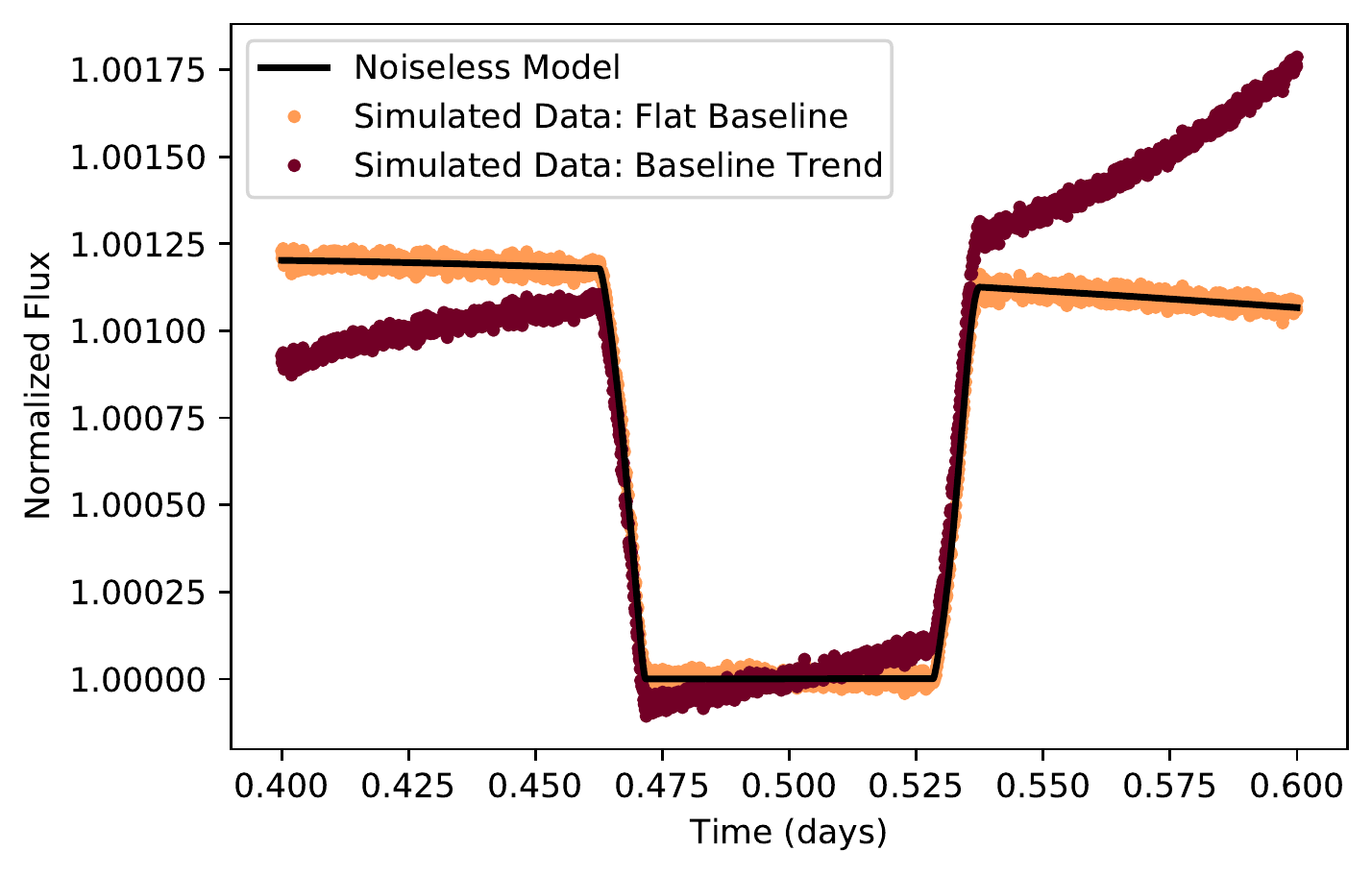}{0.48\textwidth}{Forward Model of Idealized hot Jupiter}
	\fig{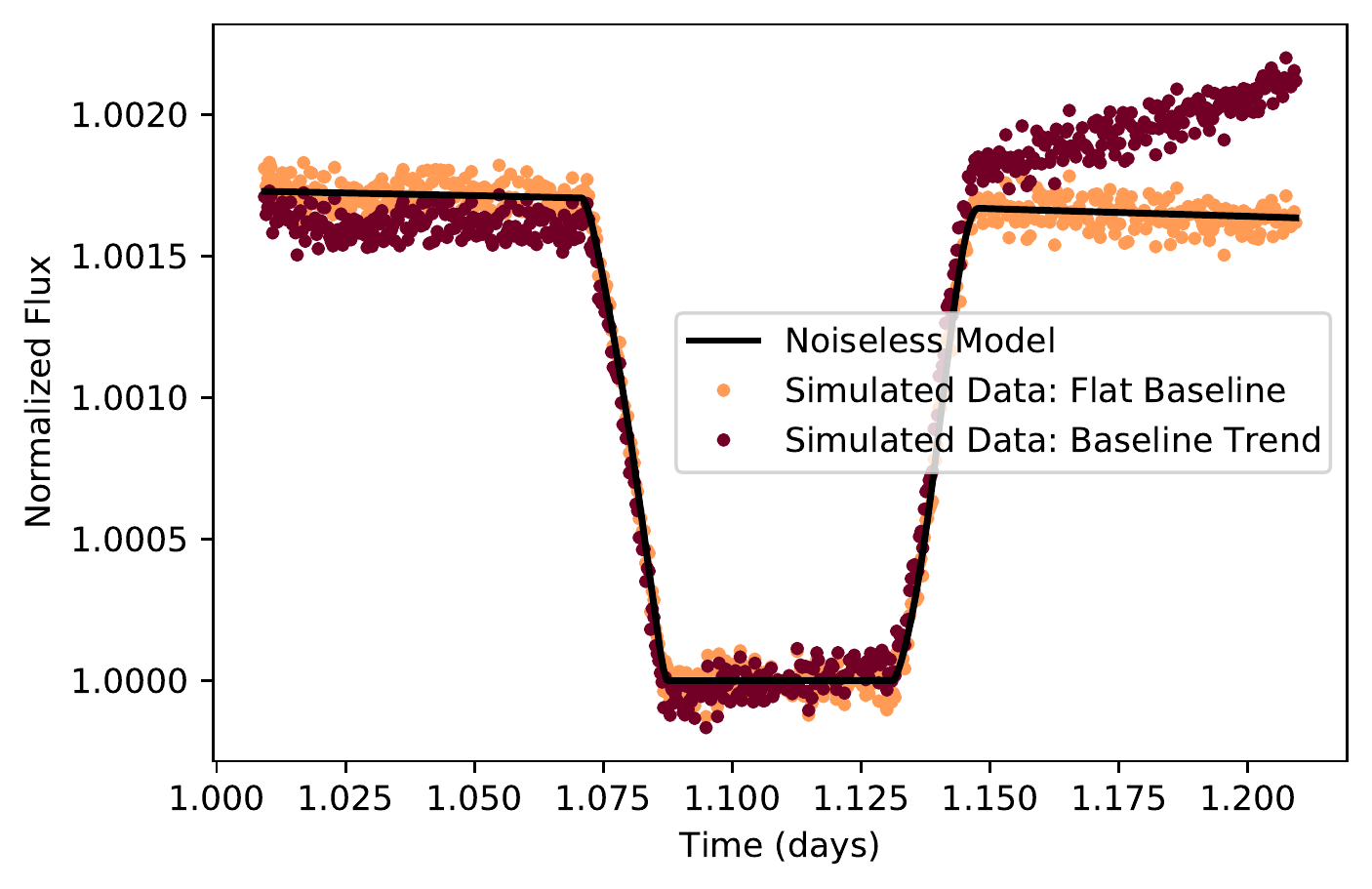}{0.48\textwidth}{Forward model of \hdb}
}
\caption{
Forward lightcurve models and simulated data with 6~ppm per minute precision for an idealized planet (left) and 31~ppm per minute precision for \hdb\ (right).
The simulated lightcurves with a flat baseline trend (other than the astrophysical variations from the rotation of the planet) are shown with light brown data points.
It is assumed that the planet is observed only near eclipse but with sufficient baseline prior to ingress and following egress for baseline characterization.
We add a cubic polynomial trend (left) and a quadratic polynomial trend (right) that could be caused by instrument artifacts or astrophysical variability (dark brown points) to understand which aspects of the recovered maps are affected by non-flat baseline trends\label{fig:forwardLC}}
\end{figure*}

\begin{figure*}
\gridline{
	\fig{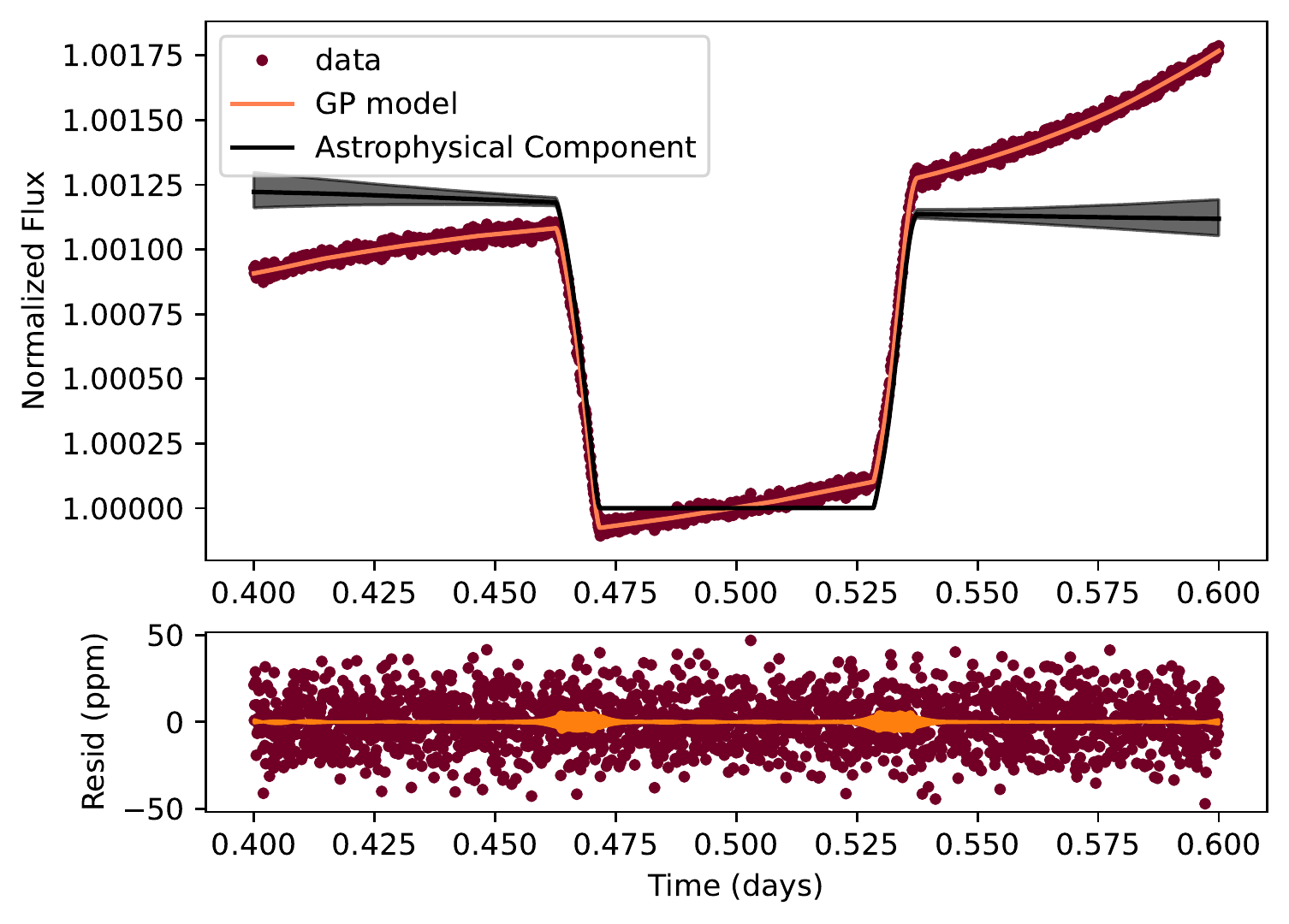}{0.48\textwidth}{Posterior Lightcurve for Idealized hot Jupiter}
	\fig{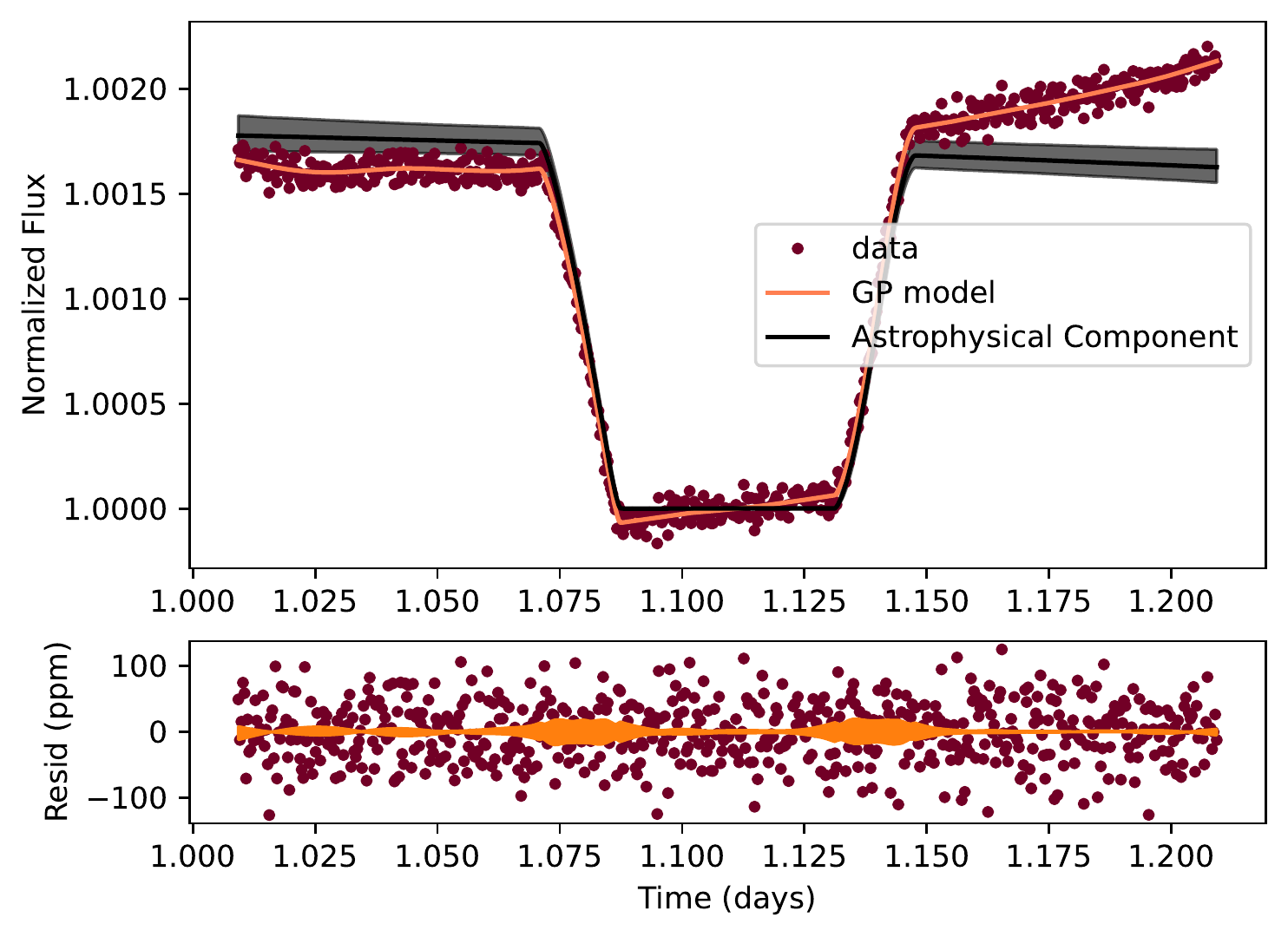}{0.48\textwidth}{Posterior Lightcurve for \hdb}
}
\caption{
Posterior lightcurves for the non-flat baselines, separated into the full Gaussian-process model (orange shaded region) and the astrophysical component (gray shaded region).
The simulated data as well as the residuals are shown in dark brown.
For both the idealized planet (left) and the JWST NIRCam HD 189733 b simulation (right), the model uncertainty grows during ingress and egress due to the map uncertainty and the fact that some mapping components only exhibit variations during eclipse.
Different maps create a range of astrophysical lightcurves and these lightcurves (encompassed in the gray region) also have a range of astrophysical baselines that grow increasingly uncertain away from mid-eclipse.
These astrophysical lightcurves can appear similar to systematic baseline trends, but are constrained more by the ingress/egress signatures of the maps. \label{fig:postLightcurves}}
\end{figure*}

\section{Forward Models}\label{sec:forwardModel}
We begin by creating an arbitrary emission map with strong asymmetries to understand how hard it is to uncover the map.
We use the map of Earth's continents fit with spherical harmonics up to degree 3, but it is hardly recognizable as the Earth at this resolution.
We chose this map instead of a more plausible hot spot Jupiter model, such as one produced by a General Circulation Model (GCM) to prepare for unexpected new maps of complicated planets.
However, we explored how the results changed with a GCM forward map, discussed in Section \ref{sec:differentForwardMap} and Appendix \ref{sec:gcmForward}.
We assume that the planet is tidally locked and on a circular orbit with a rotational period equal to the orbital period.
We also assume that the rotational axis is parallel to the orbital axis (no obliquity).
Our results are therefore applicable to aligned hot Jupiters, but misaligned planets and non-synchronized planets have been considered elsewhere in \citet{adams2021eclipseMappingRotation}.

We examine two different planet parameters:
\begin{enumerate}[noitemsep]
	\item an idealized planet at extremely high signal to noise (5.6 ppm at 1 minute cadence) and 
	\item the published parameters for \hdb. (31 ppm at 1 minute cadence) 
\end{enumerate}
For the idealized planet, we choose a Jupiter sized (0.1 R$_\odot$) planet orbiting a solar mass star with an orbital period of 1 day in a circular orbit and an impact parameter of 0.51.
A non-zero impact parameter is needed to constrain the latitude and longitude structures from two different stellar ``scans'' by the stellar limb across the planet, \citep[e.g.][]{deWit2012eclipsemap189}.
While we focus on planet inhomogeneities, we ignore any inhomogeneities in the stellar surface for the simple forward model.
For \hdb's orbital and physical parameters, we used the stellar and planet radii, period and inclination from \citet{addison2019minervaAustrialisFirstResults} arXiv version 1. We note that the final published version of the paper and arXiv version 3 had updated planet and star parameters, but these will not affect the conclusions of this work.
Table \ref{tab:forwardParams} contains a summary of the planet parameters.
The impact parameters of the idealized planet and \hdb\ ensure that there are no spherical harmonic components in the null space of the lightcurves (ie. map components that produce no lightcurve signal) \citep{cowan2013lcInclinationObliquityAlbedo,luger2021mappingStellarSurfacesDegeneracies} for up to the 3rd degree spherical harmonics.
We discuss other impact parameters in Section \ref{sec:impactParamChoice}.

We use \starry\ \citep{luger2019starry} to model the lightcurves of the forward map of the planet.
Figure \ref{fig:forwardLC} shows the lightcurves for the forward models. We assume a 4.8 hour sequence of integrations surrounding the eclipse, as would be observed by JWST.
The idealized planet lightcurve has an extremely high precision of 15 ppm per time sample of 8.4 seconds, which is 5.6 ppm per one-minute timescale.
We add noise with a Gaussian distribution that is identically distributed and independent for all data points as a starting point.
This represents a very optimistic high precision lightcurve of a future observatory to better understand the high signal-to-noise limit.
For \hdb, we simulate a NIRCam F444W long wavelength grism time series observation with the \texttt{mirage} tool \citep{hilbert2019mirage}.
We assume the same exposure parameters as GTO program 1185, which uses the Subgrism64 subarray, 4 output channels, 1 exposure and a BRIGHT2 pattern with 4 groups per integration for 6533 integrations.
We then use the uncalibrated data from the simulation and run it through the \texttt{jwst} pipeline\footnote{\url{https://jwst-pipeline.readthedocs.io/en/latest/index.html}} with a custom extraction using \texttt{tshirt}\footnote{\url{https://tshirt.readthedocs.io/en/latest/}} \citep[e.g.][]{glidic2022corot1}.
We combined all wavelengths into a broadband lightcurve and use the robust standard deviation of the lightcurve to create simulated Gaussian noise to the lightcurve, which discards cosmic ray outliers.
The resulting robust estimator for the standard deviation is 145 ppm at a cadence of 2.725 seconds, which is 31 ppm per minute.
We bin the simulated lightcurves for \hdb\ to a cadence of 27 seconds at a precision of 45 ppm for better visualization of the eclipse lightcurves.

Next, we create arbitrary polynomial baselines to add to the lightcurve to represent an unknown astrophysical or instrumental trend.
The polynomials are multiplied by the flux to create a non-flat baseline, as visible in Figure \ref{fig:forwardLC} (dark brown curves).
We save the lightcurves both before and after adding the polynomial baselines so that we can directly see the effect of a baseline trend on the final map interpretation of a planet.
For the idealized planet, we use a 3rd order baseline with coefficients of 0.14\%/hr, 0.46\%/hr$^2$ and 2.21\%/hr$^3$ and for \hdb\ use a 2nd order baseline with coefficients of 0.14\%/hr and 0.46\%/hr$^2$.
These were arbitrarily chosen coefficients before JWST launch to have a trend that is significant compared to the eclipse depth but does not dominate the eclipse signal.
These coefficients are within a factor of 2 from preliminary analysis of JWST F444W lightcurves of \hdb\ in GTO program 1185, which have coefficients of -0.3\%/hr and 0.9\%/hr$^2$ for the baseline trend.
We also tested a 3rd order baseline and a Gaussian process for \hdb\ to see if it affected our conclusions.

\begin{figure*}
\gridline{
	\fig{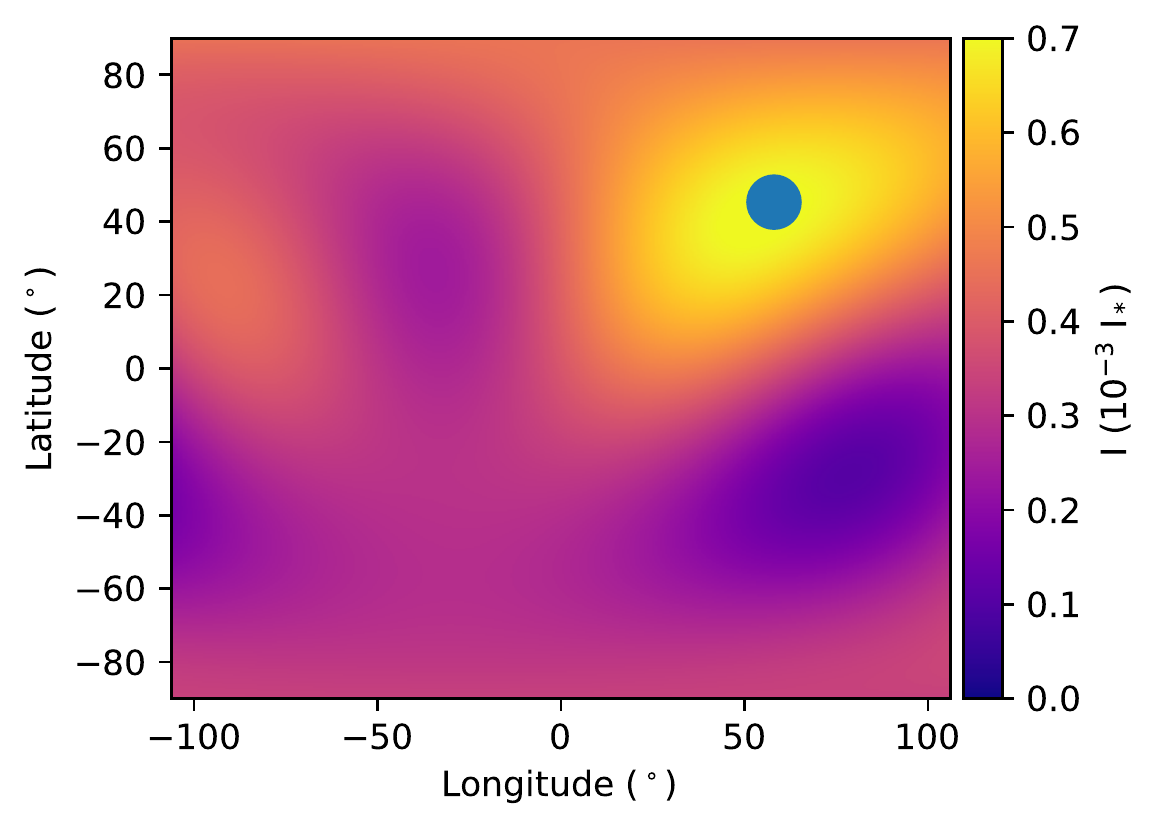}{0.3\textwidth}{Forward Input Map}
	}
\gridline{
	\fig{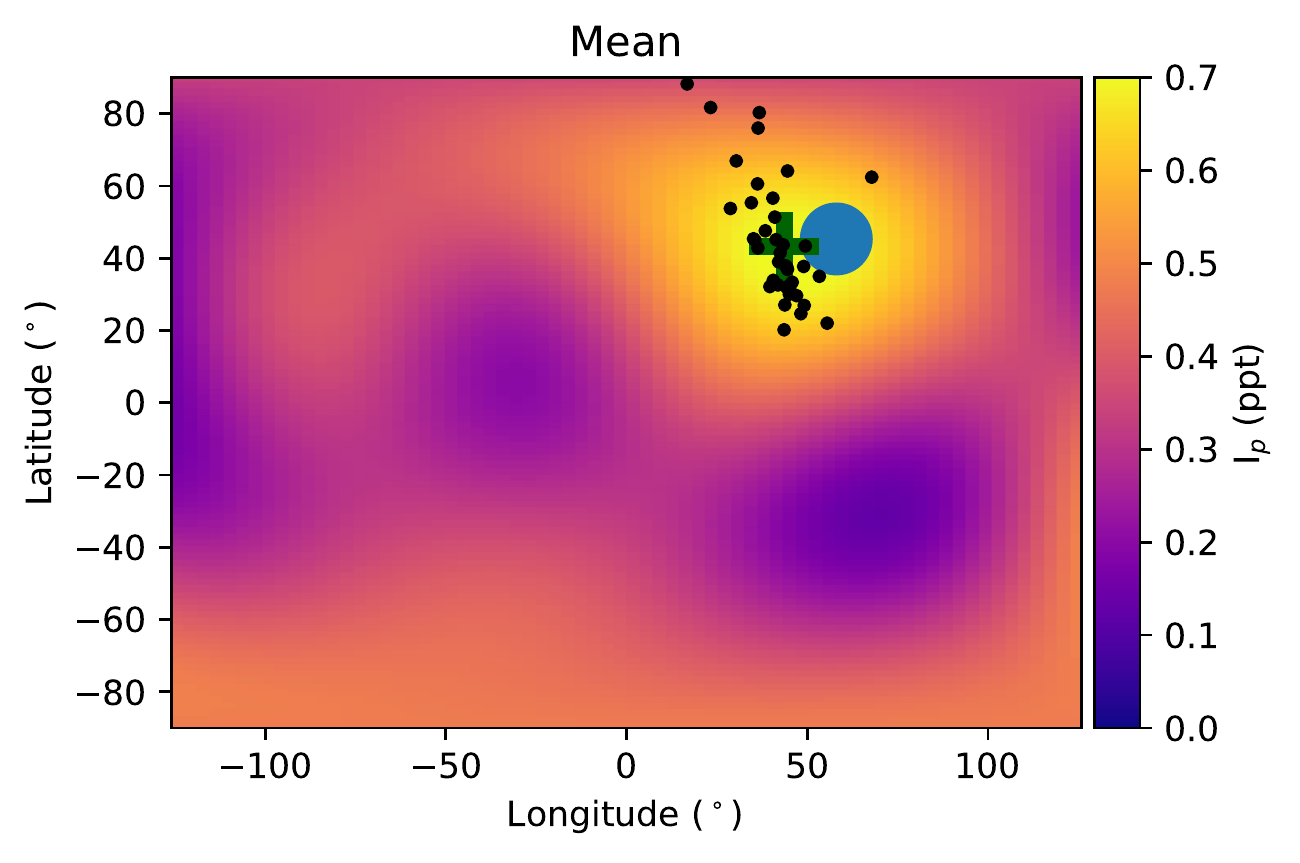}{0.30\textwidth}{Recovered \starry\ Mean with Flat Baseline}
	\fig{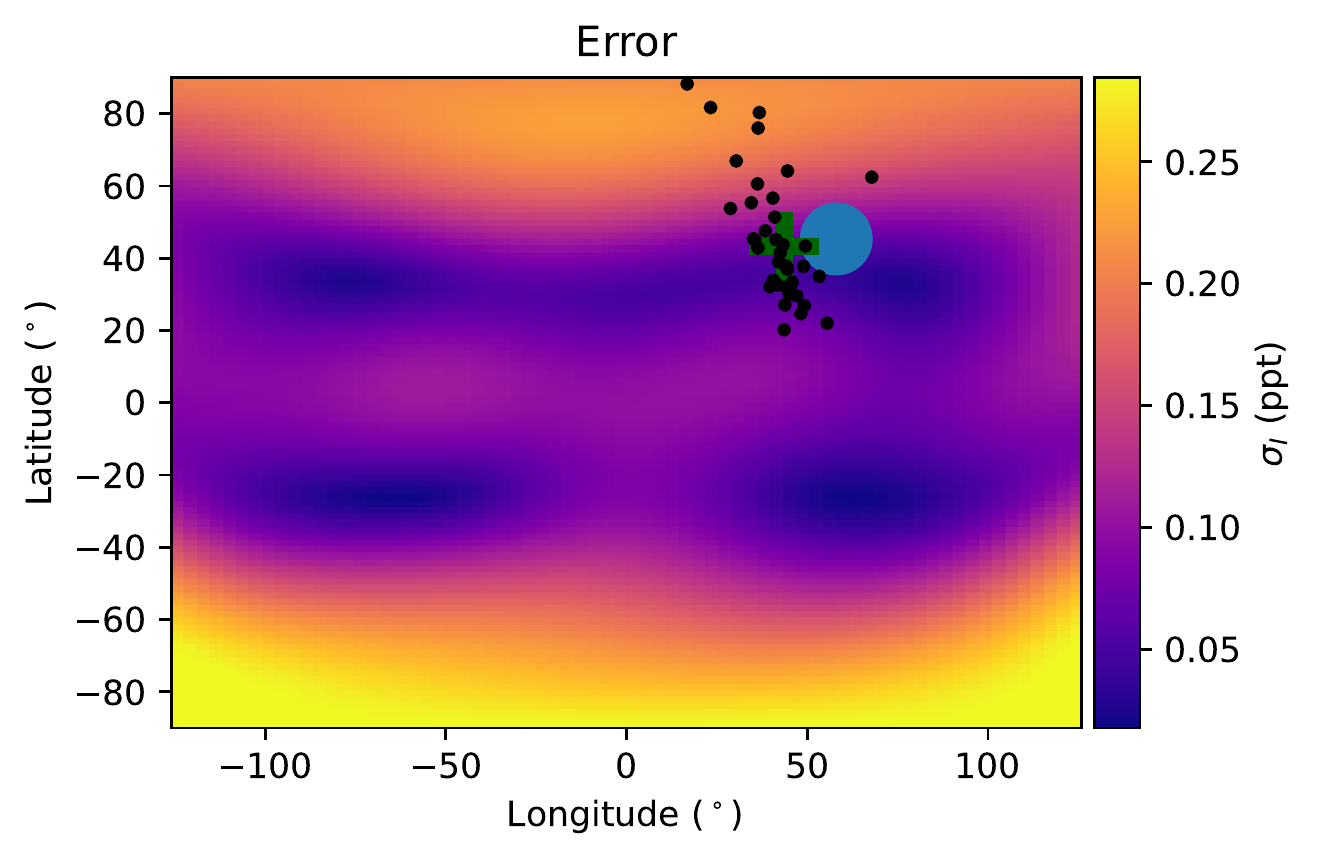}{0.30\textwidth}{Recovered \starry\ Uncertainty with Flat Baseline}
	\fig{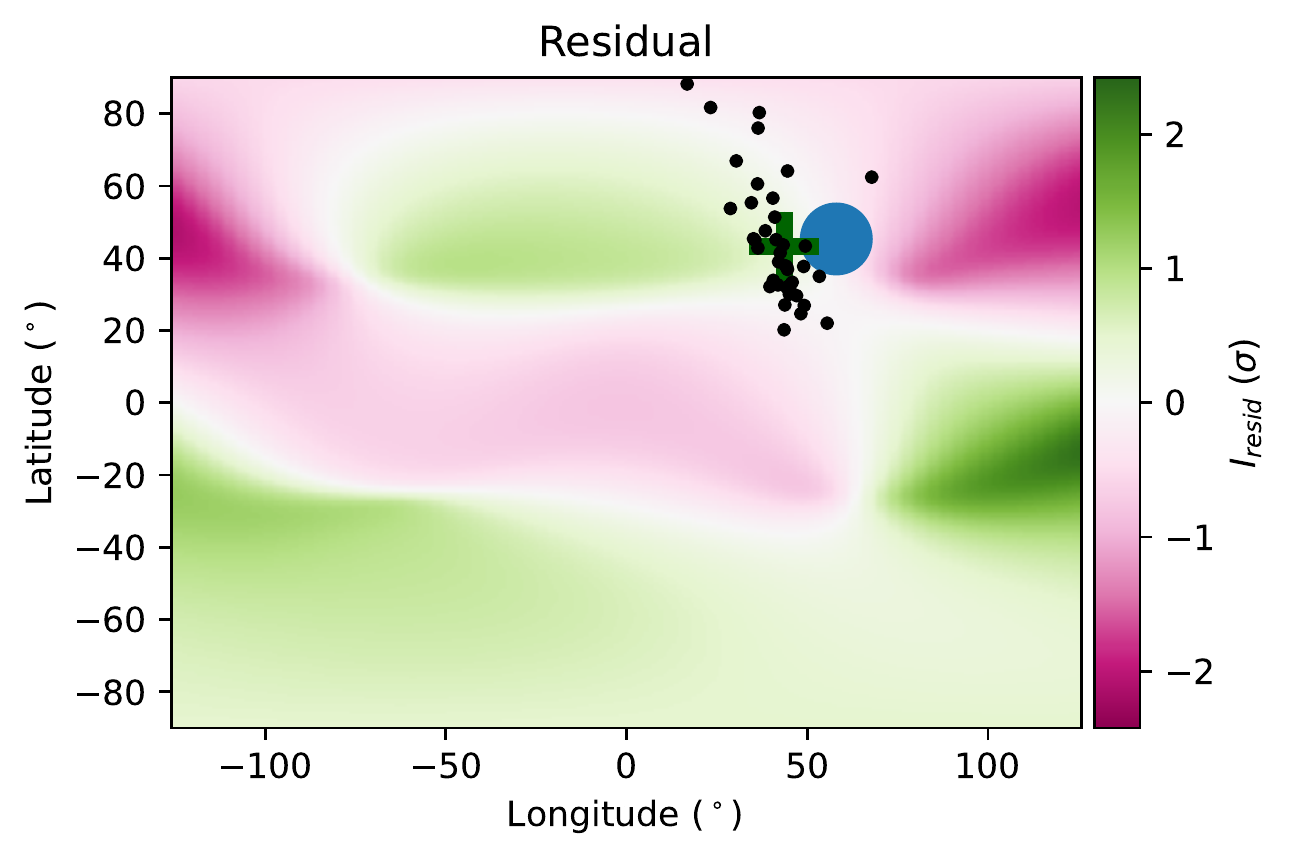}{0.30\textwidth}{Residuals with Flat Baseline}
	}
\gridline{
	\fig{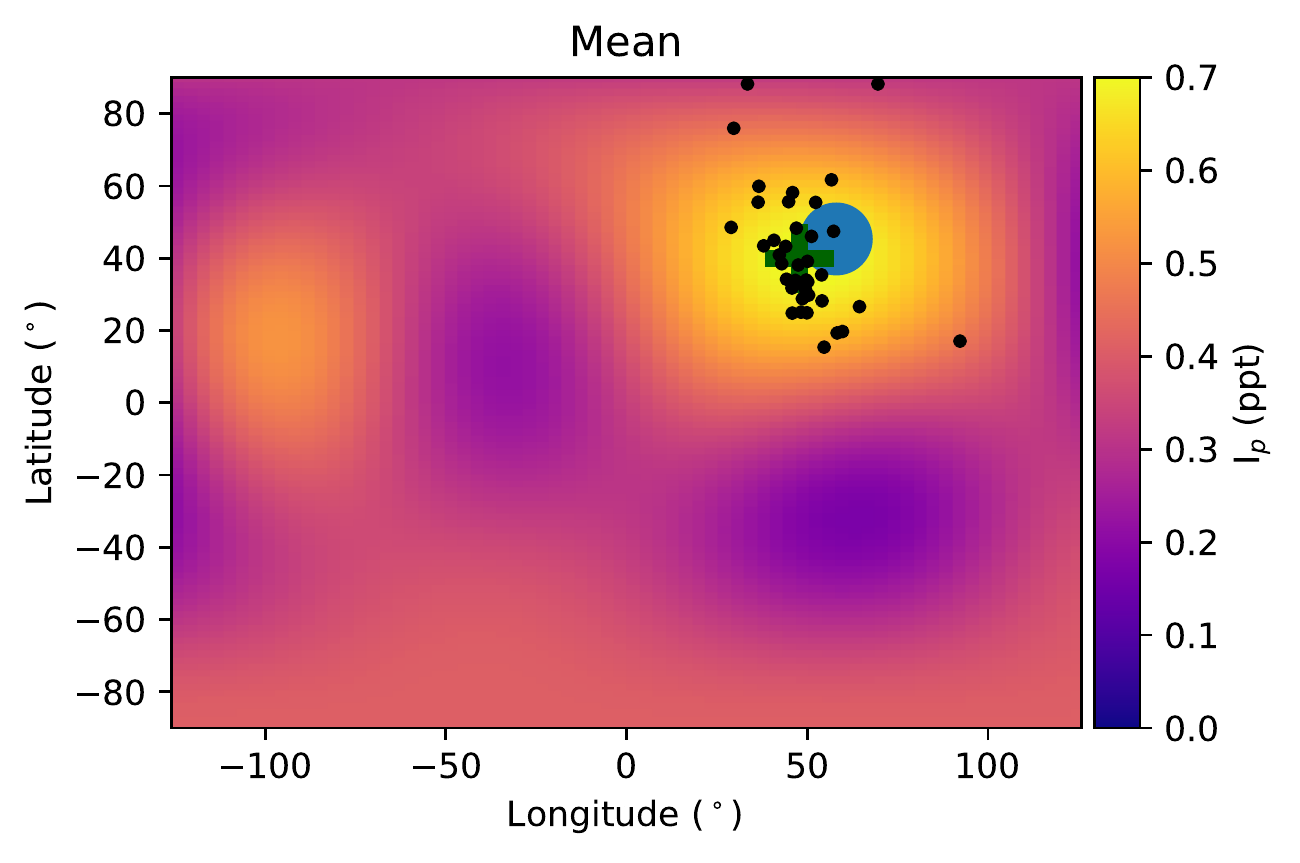}{0.30\textwidth}{Recovered \starry\ Mean with Cubic Baseline}
	\fig{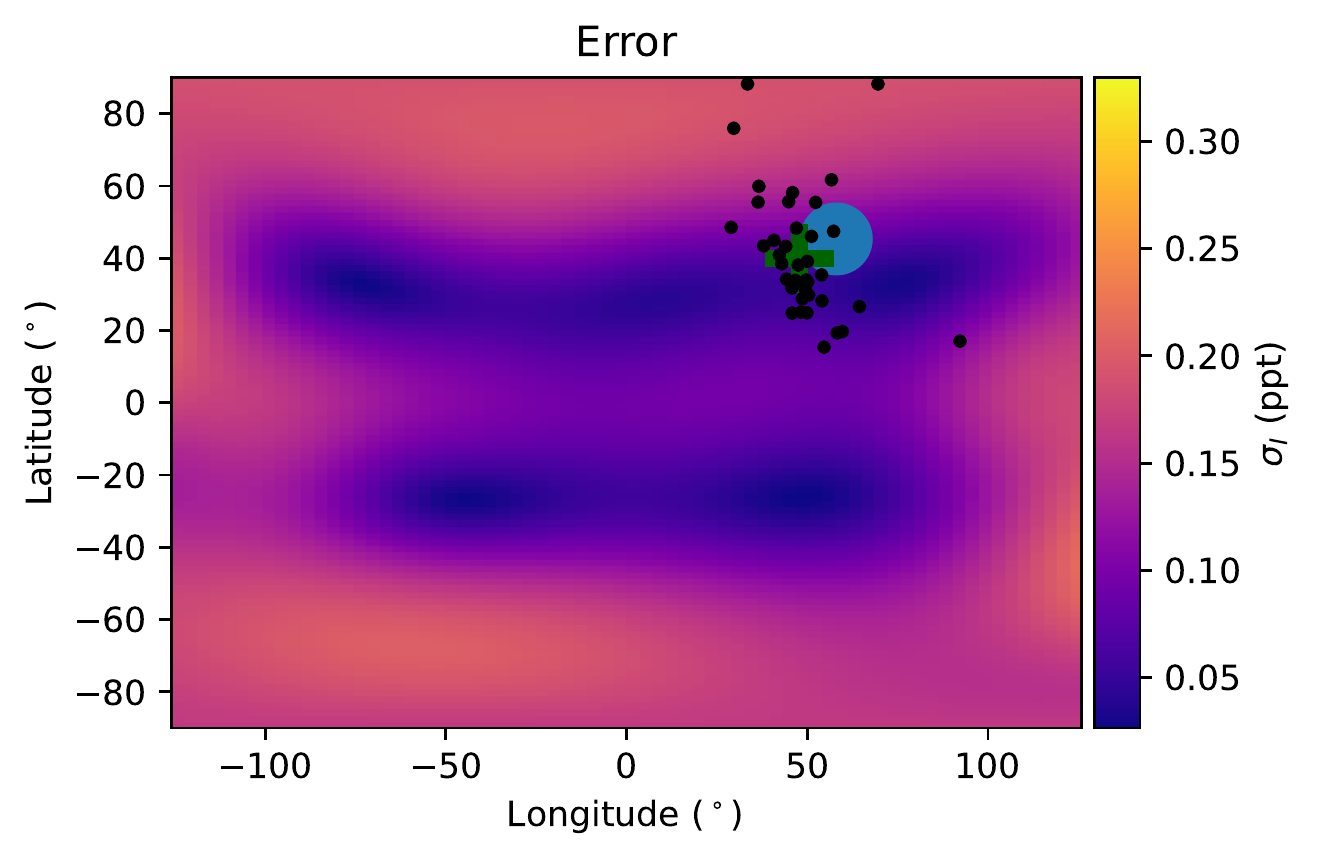}{0.30\textwidth}{Recovered \starry\ Uncertainty with Cubic Baseline}
	\fig{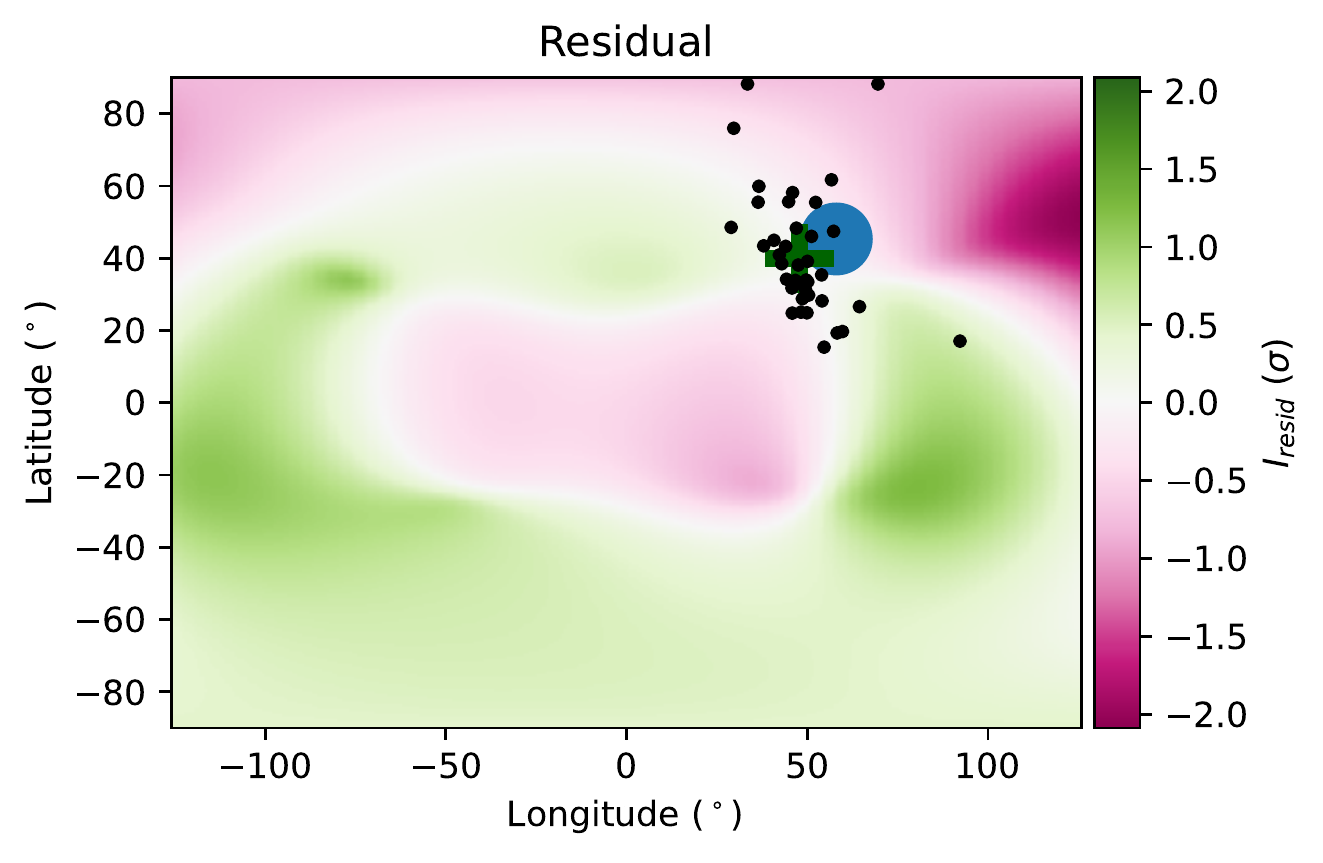}{0.30\textwidth}{Residuals with Cubic Baseline}
	}
\caption{{\it Top:} Arbitrary map model for the hot Jupiter planet, which is actually from an Earth map up to spherical harmonic degree 3 just to provide complicated structures.
The location of peak brightness is shown with a blue circle.
{\it Middle:} \starry\ map mean, uncertainty and residuals for the idealized planet where the baseline is flat
{\it Bottom:} \starry\ map mean, uncertainty and residuals for the idealized planet where the lightcurve has a cubic polynomial trend. 
Black points show the locations of peak brightness for individual map samples drawn from the posterior and the green plus shows the peak brightness Gaussian fit centroid to the mean map \label{fig:starryMapFitsIdealized}.
The added baseline does not significantly affect the mean or error map, but does affect the precision of the peak brightness location as well as the residuals.}
\end{figure*}

\section{Recovered Maps}\label{sec:recoveredMap}

\subsection{Methodology of Lightcurve Fitting with Spherical Harmonics}\label{sec:sphHarmonicFits}
We fit the lightcurves for the simulated data with \starry\ and with spherical harmonics up to 2nd or 3rd degree using the \texttt{pymc3} \citep{salvatier2015probProgramming} probabilistic programming suite.
First, we fit the forward model that has a flat baseline (light brown data points in Figure \ref{fig:forwardLC}).
We assume that the planet orbit is known perfectly and fix the orbital parameters at the same values as the forward model to focus on the mapping parameters.
We attempted to fit the lightcurves with both with a 2nd degree and 3rd degree spherical harmonic map and choose the lower value of the Bayesian Information Criterion (BIC) to choose between the two models.
For the idealized planet, this was 3rd degree (up to $\ell=3$) and for \hdb, it was up to 2nd degree (up to $\ell=2$).
The 3rd degree model has 16 free mapping parameters because there are 16 spherical harmonics terms at 3rd degree.
The 2nd degree model has 9 free mapping parameters for 9 terms.
\starry\ scales all spherical harmonics by an overall amplitude term (\texttt{amp}), but it is degenerate with $Y_{0,0}$, so the value for $Y_{0,0}$ is fixed at 1.0, maintaining 16 free mapping parameters for the 3rd degree spherical harmonics and 9 free mapping parameters for the 2nd degree spherical harmonics.
There is additionally one free parameter for the standard deviation of the lightcurve and, in some models, 2 more Gaussian process parameters, depending on whether one assumes that the lightcurve baseline is flat or not.

We experimented with different map priors to understand the best way to sample a physical map:
\begin{enumerate}
\item direct spherical harmonic sampling with a positivity constraint and
\item pixel sampling and transformation to spherical harmonics.
\end{enumerate}
In our first approach, we assumed Gaussian independent priors for the spherical harmonic coefficients with a mean of 0.05 and a standard deviation of 0.5.
These are normalized by the amplitude, which has a prior of 0.17$\pm 0.05 \%$ for the idealized planet and 0.10 $\pm 0.02 \%$ for \hdb.
Significant portions of the multidimensional parameter space of these spherical harmonic priors consist of unphysical models, where the flux goes negative in some parts of the map and is compensated by the amplitude and other spherical harmonic coefficients to ensure the overall eclipse depth is correct.
To ensure positivity, we use the $\texttt{pymc3 Potential}$ function to force the map to be non-negative by evaluating the map at a resolution of 100 by 100 pixels on a rectangular grid and setting the probability to negative infinity if any pixels at visible longitudes (-106.6 to 106.2 degrees) have a flux less than zero.
In our second approach, we used \starry\ matrix transformations of the 3rd degree spherical harmonics to 30 map pixels across the map's globe.
This matrix transformation allows us to set priors on the pixels and sample them as random variables, transform these into spherical harmonics and calculate lightcurves.
We set uniform priors on all pixels from 0.0 to 1.0 times a normalization factor.
The normalization factor was given the same prior as the amplitude.
We find that the two approaches of direct spherical harmonic sampling and pixel sampling give very similar results, but the pixel sampling showed larger biases away from the input truth values (even when doing an apples-to-apples comparison where both pixel sampling and the negative infinity potential function both require non-negative maps over the entire globe).
We therefore proceed with the first approach of direct spherical harmonic sampling with a positivity constraint over the visible longitudes.

To fit the polynomial baselines, we used a Gaussian Process (GP) regression with the \texttt{celerite2} code \citep{foreman-mackey2018celerite}, which allows a flexible range of GP models.
We assume that the GP kernel follows a stochastically driven harmonic oscillator function, which allows both periodic and damping terms to model a wide range of data correlations.
This is captured as a Stochastically driven Harmonic Oscillator Term (SHOT) kernel function in \texttt{celerite2} with a highly dampened oscillator that has a quality factor (Q) of 0.25.
This fit has the same 16 free mapping parameters as the flat model and a variable for the lightcurve error value, but also includes free parameters for the Gaussian process standard deviation and the Gaussian process damping time.

\begin{figure*}
\gridline{	\fig{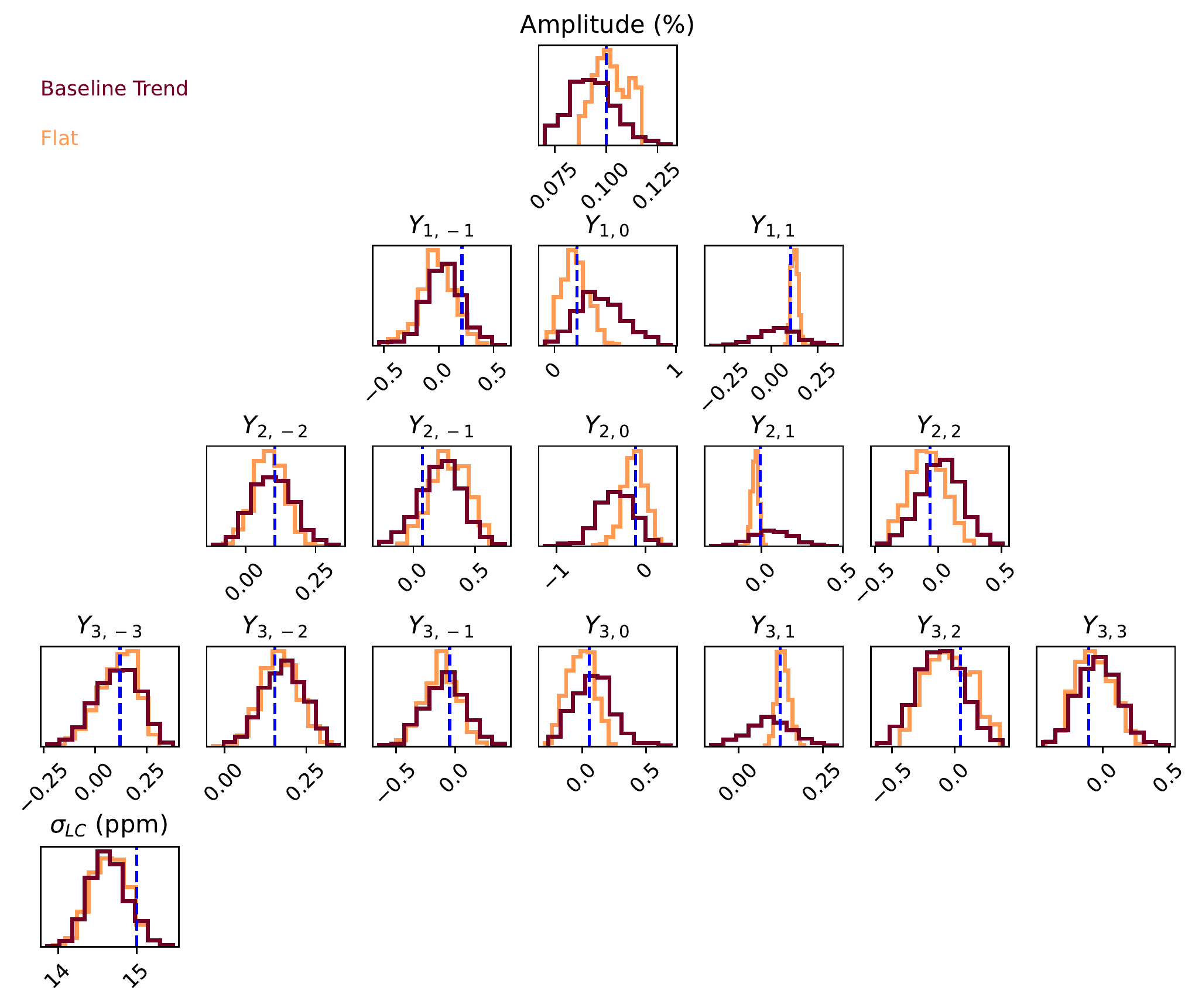}{0.99\textwidth}{Posterior Distributions for a Ultra-High Precision Planet Observation}
	}
\caption{Posterior distributions for all mapping variables and the lightcurve standard deviation.
The posteriors for a flat baseline are shown in light brown, the posteriors with a baseline trend are shown in dark brown, and the inputs for the forward model are shown as blue vertical dashed lines.
The most-affected parameters are the m=1 spherical harmonics.
The amplitude, shown in the top, is the integrated flux of the planet over 4$\pi$ compared to the integrated flux of the star, with units in percentage.
The lightcurve error is in units of ppm and the spherical harmonic coefficients are unitless because they are ratios to the $Y_{0,0}$ term.
\label{fig:posteriorHistIdealP}}
\end{figure*}

\begin{figure*}
\gridline{
	\fig{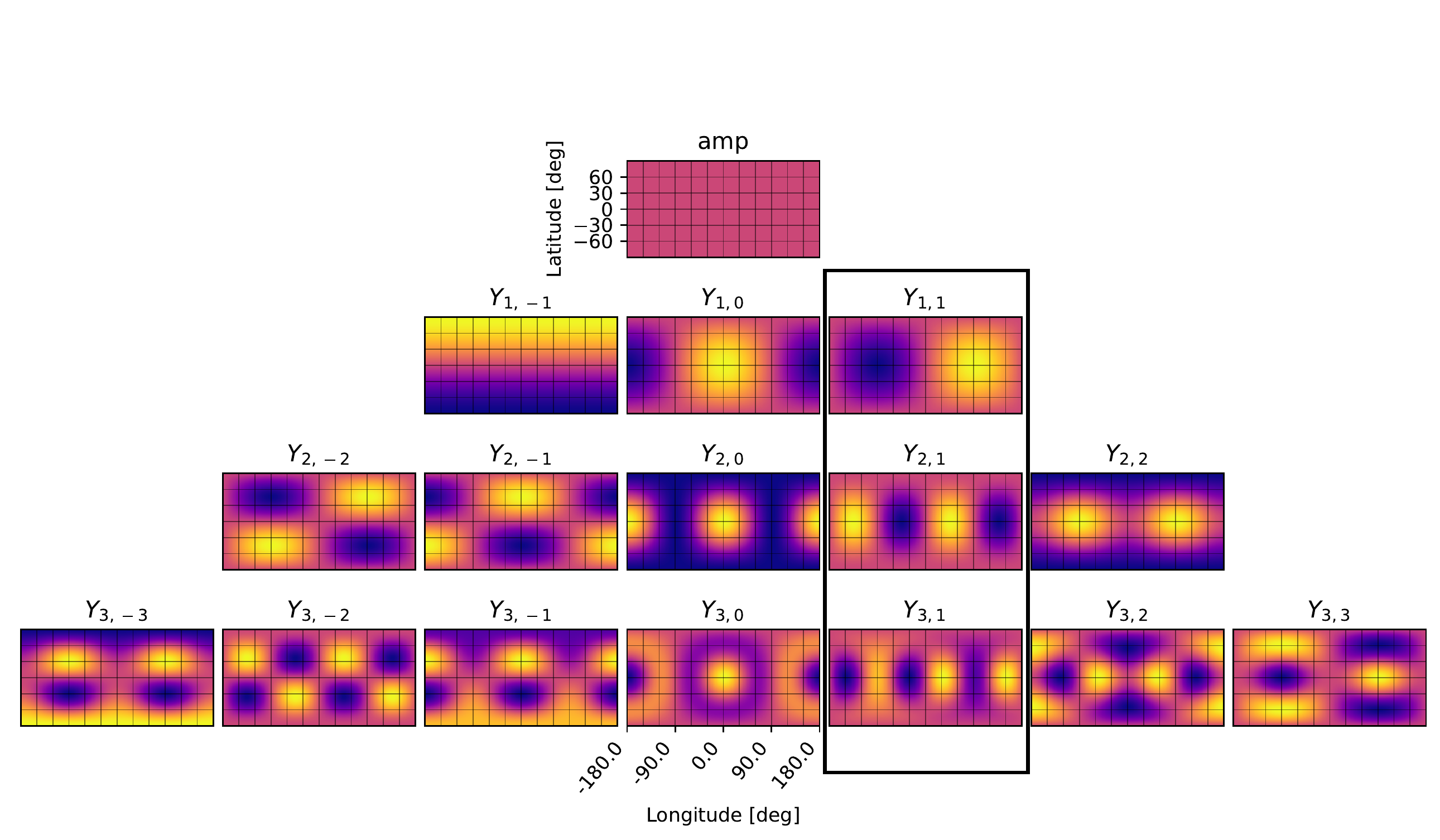}{0.99\textwidth}{Maps of Spherical Harmonics Through $\ell$=3}
	}
\gridline{
	\fig{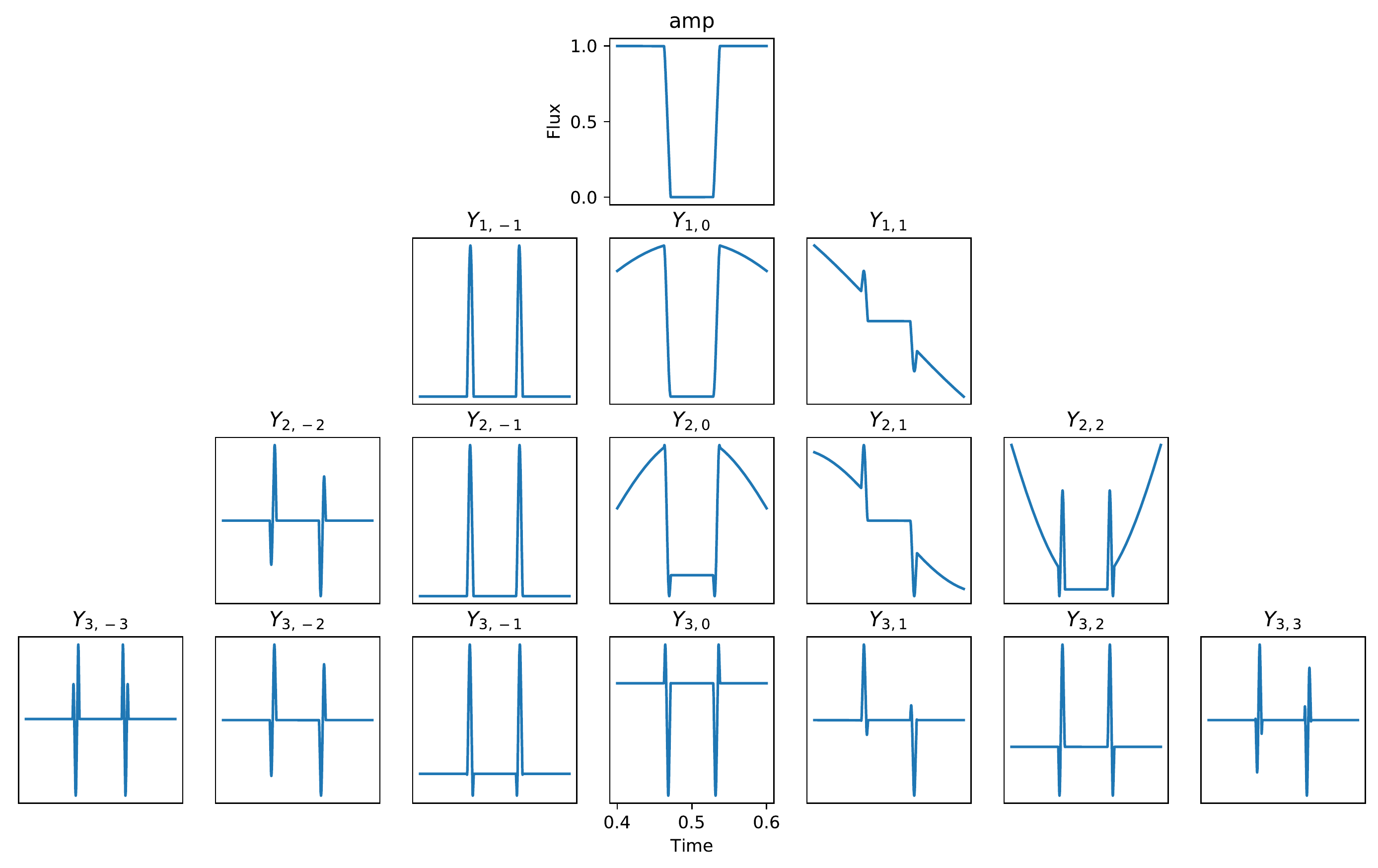}{0.79\textwidth}{Spherical Harmonic Lightcurves}
	}
\caption{The spherical harmonic maps (top) and lightcurves (bottom) are shown up to degree 3.
The 3 terms most affected by baseline trends (Y$_{1,1}$, Y$_{2,1}$ and Y$_{3,1}$) are highlighted.
	The baseline primarily affects the horizontal distribution of the planet.
	While the Y$_{1,1}$ and Y$_{2,1}$ terms have strong linear baselines, the Y$_{3,1}$ spherical harmonic lightcurve correlates strongly with Y$_{1,1}$ and $Y_{2,1}$ as shown in Figure \ref{fig:posteriorCorner} \label{fig:sphHarmonicLCs}.
	}
\end{figure*} 

\subsection{Spherical Harmonic Fitting Results - Idealized Planet}
The fitted lightcurves, separated into the astrophysical and Gaussian process components, are shown in Figure \ref{fig:postLightcurves}.
The forward map as well as the resulting \starry\ posterior map distributions are shown in Figure \ref{fig:starryMapFitsIdealized}.
The \starry\ fits for the idealized planet recover the main bright feature at the Northeastern part of the forward map.
The mid-latitudes have the lowest uncertainty and it grows slightly toward the equator and more dramatically at the planet poles.
In both the flat baseline and the curved baselines, the Northern and Southern poles have the largest uncertainty, due to the smaller projected impact on the secondary eclipse.
We note that the mean maps in Figure \ref{fig:starryMapFitsIdealized} are very similar whether the lightcurve has a flat baseline and no Gaussian process fitting and a random cubic baseline shown in Figure \ref{fig:forwardLC} (left).
However, we will see that some longitudinal information is lost from the baseline uncertainty.
The residual maps are shown in the right plots of Figure \ref{fig:starryMapFitsIdealized} where residuals is calculated as $r(x,y) = (\bar{f}(x,y) - t(x,y))/\sigma(x,y)$, where $r(x,y)$ is the residual for a given latitude $x$ and longitude $y$, $\bar{f}(x,y)$ is the mean map from the posterior, $t(x,y)$ is the true input map and $\sigma(x,y)$ is the error in the posterior map.

We next study the location of the location of peak brightness in both the forward and recovered maps.
For the peak brightness fitting, we evaluate maps in a rectangular projection and model the peak with a two dimensional Gaussian function and find the least-squares fit.
The forward map's location of peak brightness is located at 59$\degree$ longitude East of the subsolar point and 48$\degree$ North of the subsolar point.
In Figure \ref{fig:starryMapFitsIdealized}, we show the location of peak brightness of the forward model (blue circle), the location of peak brightness of random maps drawn from the posterior (black points) and the location of peak brightness of the mean map (plus symbol).
While the error maps for the flat baseline and the cubic baseline appear very similar, the recovered distribution of peak brightness location appears different for the two kinds of baselines.

Figure \ref{fig:starryMapFitsIdealized} shows that the flat baseline results in a narrower range of recovered peak brightness locations than when there is greater baseline uncertainty.
The individual draws are clustered together more closely in longitude and latitude.
However, this cluster of points is offset from the forward input map and has a long tail of possible values extending through the true input input brightness peak.

Another way to examine the map features is the posterior distributions of the spherical harmonic coefficients shown in Figure \ref{fig:posteriorHistIdealP}. Figure \ref{fig:posteriorHistIdealP} includes all of the spherical harmonic term variables (other than $Y_{0,0}$, which is fixed at 1.0), the map amplitude and the standard deviation variable for the lightcurve error model.
As expected, all of the posteriors correctly recover the input true variables of the forward model (dashed blue vertical lines).
While not apparent from the mean maps and errors, there are significantly different precisions for the $Y_{1,1}$, $Y_{2,1}$ and Y$_{3,1}$ spherical harmonics depending on the assumptions and existence of a baseline trend.
As seen in Figure \ref{fig:sphHarmonicLCs}, the $Y_{1,1}$, $Y_{2,1}$ and Y$_{3,1}$ terms describe the longitudinal distribution of the map, as seen by their individual maps.
The posterior distributions for the $Y_{1,1}$, $Y_{2,1}$ and Y$_{3,1}$ coefficients are significantly affected by the injected baseline trend and its associated uncertainty, reducing the 68\% posterior interval (ie precision) by a factor of 5.8, 5.4 and 2.8 respectively.
The amplitude and the $Y_{2,0}$ term are also affected at a smaller level, reducing the precision by a factor of 1.4.

A closer inspection of the $Y_{1,1}$ and $Y_{2,1}$ spherical harmonic terms, as shown in Figure \ref{fig:sphHarmonicLCs}, reveals that they contain a significant slope to the phase curve (ie. baseline before and after eclipse).
This is intuitive from the spherical harmonic maps showing the longitudinal dependence of the lightcurves, also shown in Figure \ref{fig:sphHarmonicLCs}.
If there is a systematic slope in flux near eclipse due to an instrument or starspot/facula on the host star, it is clear that information about the $Y_{1,1}$ and $Y_{2,1}$ terms will be highly correlated with the slope.
Similarly, if there is a curvature in the baseline, it can decrease the precision of the $Y_{1,0}$ and $Y_{2,0}$ spherical harmonic terms because they contain significant curvature out of eclipse.
What is less clear is the $Y_{3,1}$ term, which has no variation outside of eclipse.
The $Y_{3,1}$ term's lightcurve, however, has a similar peak at ingress and valley at egress as the $Y_{1,1}$ and $Y_{2,1}$ spherical harmonic terms.
Thus, the $Y_{3,1}$ lightcurve correlates significantly with the $Y_{1,1}$ and $Y_{2,1}$ lightcurves.
This manifests itself in the correlated random variables for the coefficients of the Y$_{1,1}$, $Y_{2,1}$ and $Y_{3,1}$ spherical harmonics.

\begin{figure*}
\gridline{
	\fig{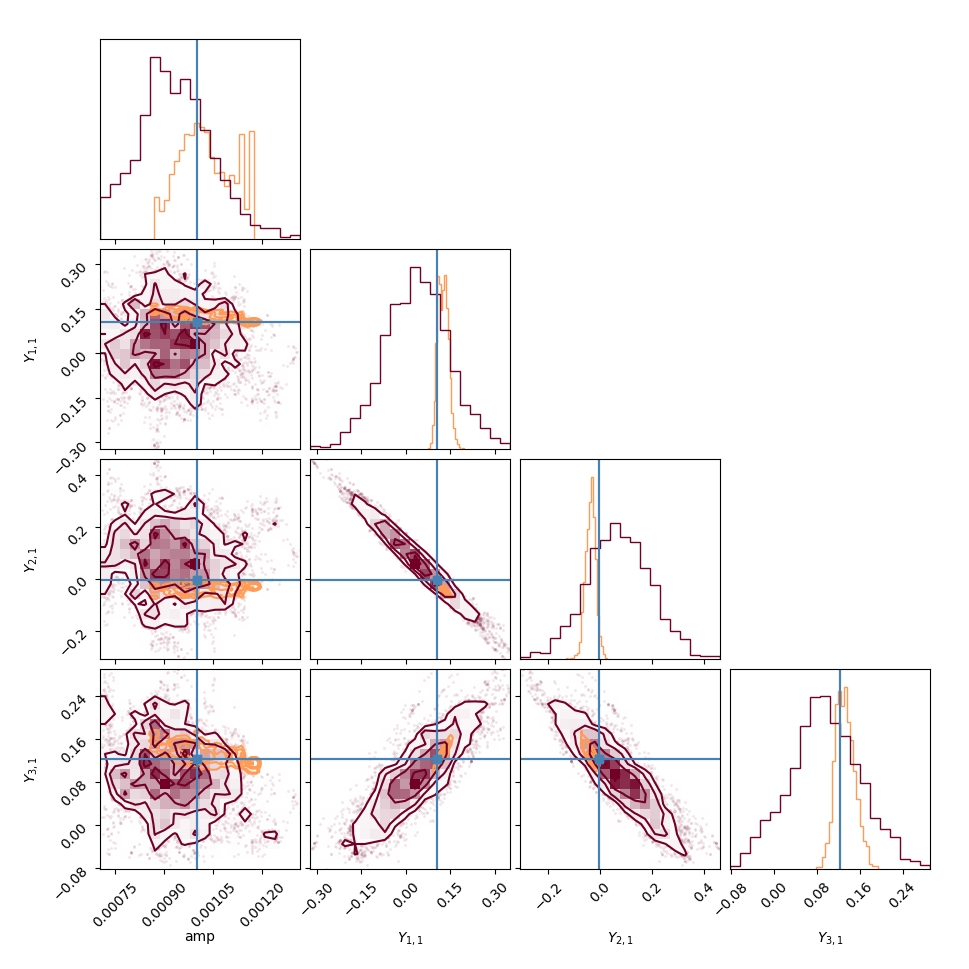}{0.99\textwidth}{Posterior of Selected Variables}
	}
\caption{Posterior distributions for the four model parameters most affected by the baseline trend for the idealized planet lightcurve shown in Figure \ref{fig:forwardLC}.
The $Y_{1,1}$, $Y_{2,1}$ and $Y_{3,1}$ terms are highly correlated, so uncertainty in $Y_{1,1}$ and $Y_{2,1}$ from the baseline trend also increases the uncertainty in $Y_{3,1}$ for the case where there is a baseline trend (dark brown).
A lightcurve with no baseline trend and pure astrophysical signal (light brown posteriors) gives much tighter constraints \label{fig:posteriorCorner}}
\end{figure*}

As shown in Figure \ref{fig:posteriorCorner}, there is indeed a strong correlation between the $Y_{1,1}$ and $Y_{2,1}$ and Y$_{3,1}$ spherical harmonic map coefficients.
These correlations can create a larger uncertainty in the $Y_{3,1}$ component even though it has no flux variation outside of eclipse.
Another consequence of these correlations is that when one variable has a bias away from the truth, the others do as well: Y$_{1,1}$'s posterior median is lower than the truth, Y$_{2,1}$'s posterior median is higher than the truth (anti-correlated) and $Y_{3,1}$'s posterior median is lower than the truth (correlated), but all are within the 68\% highest probability density interval.
An alternative parameterization of orthogonal lightcurve vectors \citep{rauscher2018moreInformativeMapping,challener2021ThERESA} could more directly reveal which map components will be affected by flux trends in the data.
We use this method in Section \ref{sec:eignmapFits}.

\begin{figure*}
\gridline{
	\fig{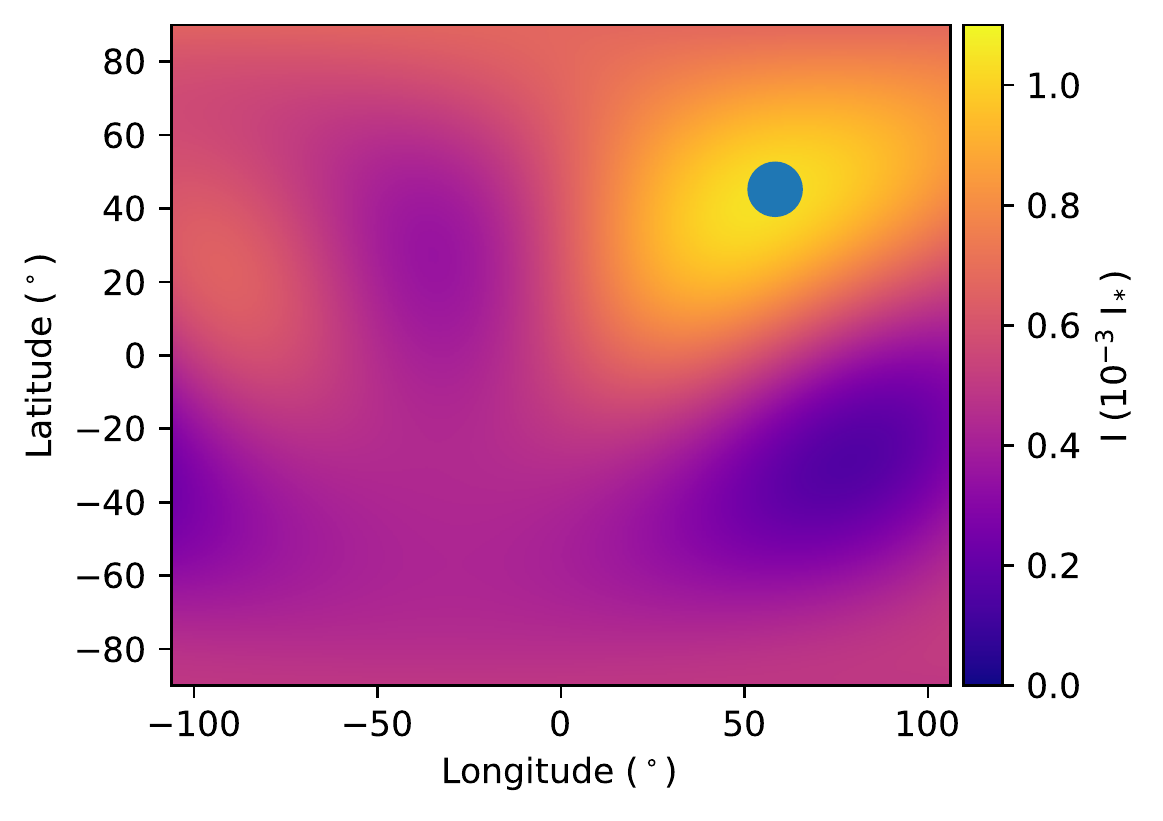}{0.3\textwidth}{Forward Input Map}
	}
\gridline{
	\fig{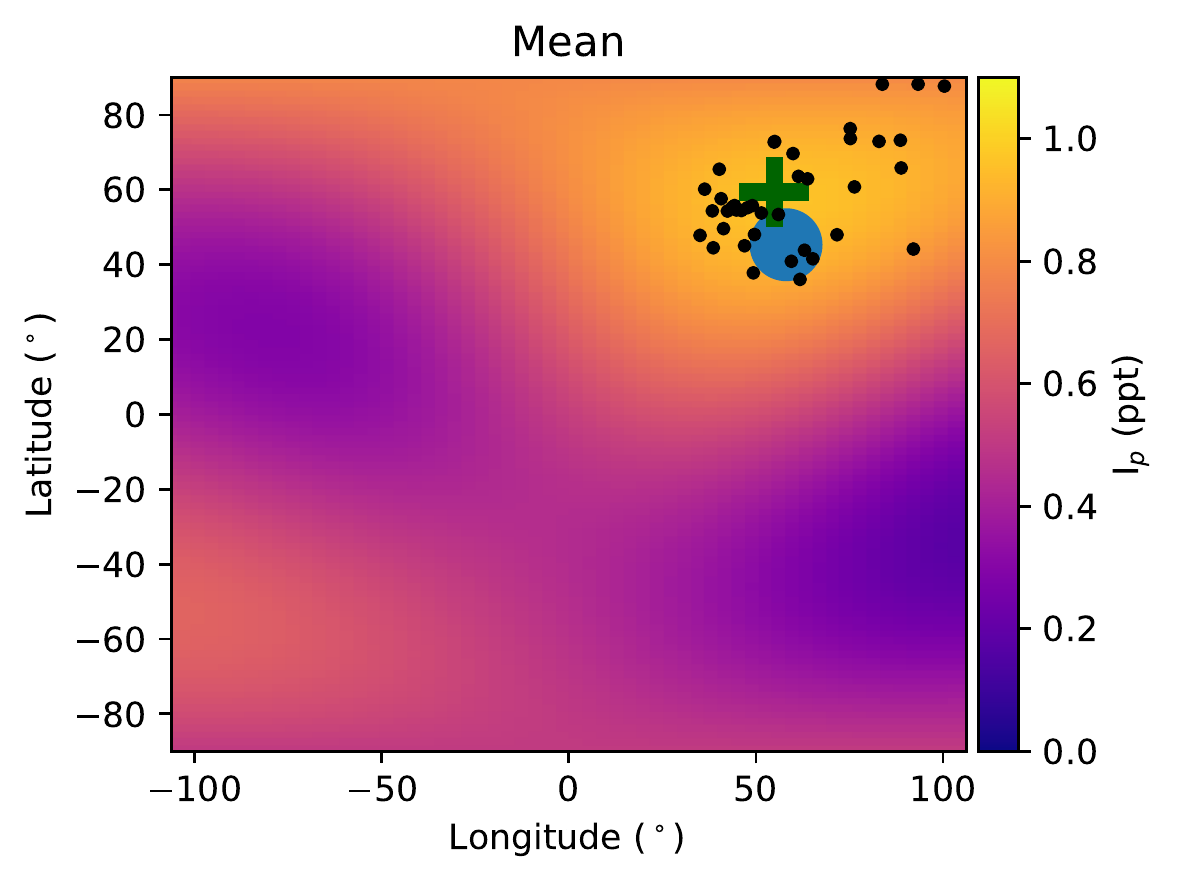}{0.30\textwidth}{Recovered \starry\ Mean with Flat Baseline}
	\fig{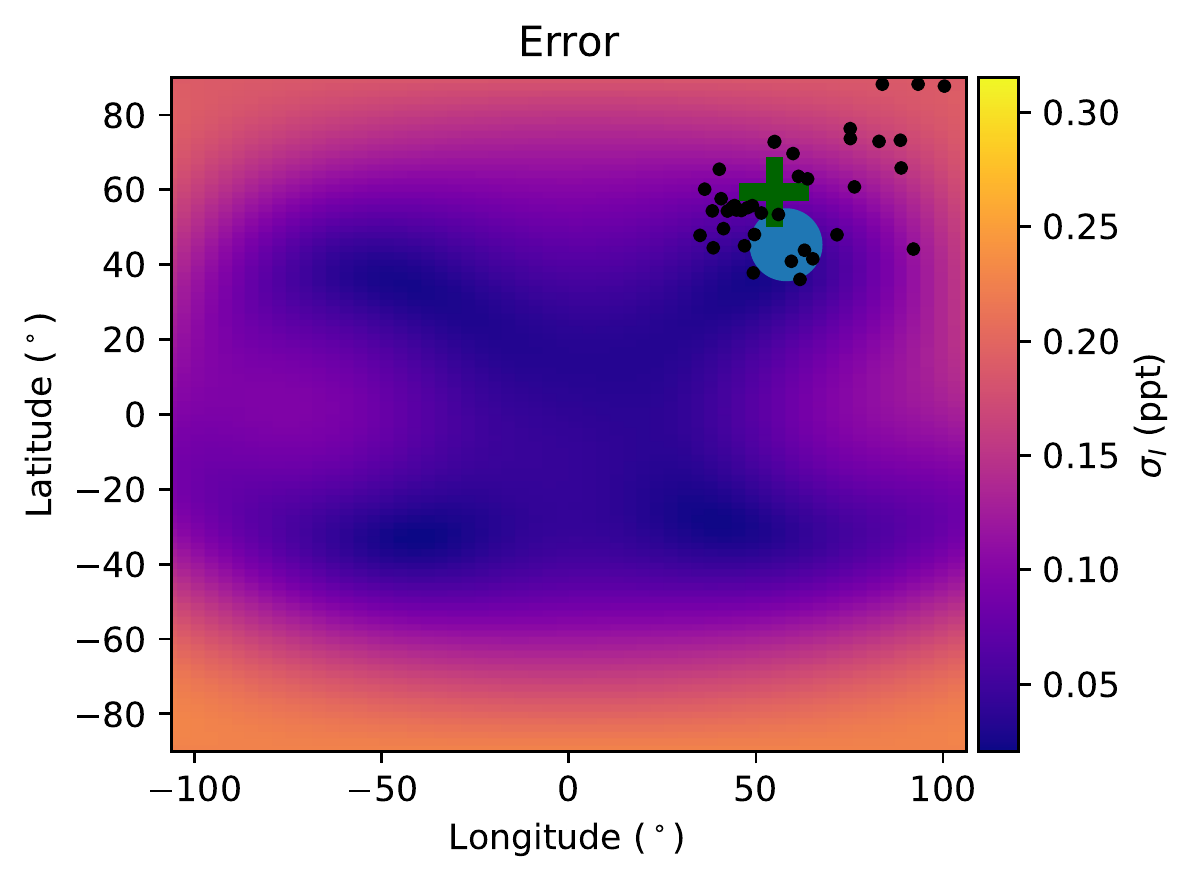}{0.30\textwidth}{Recovered \starry\ Uncertainty with Flat Baseline}
	\fig{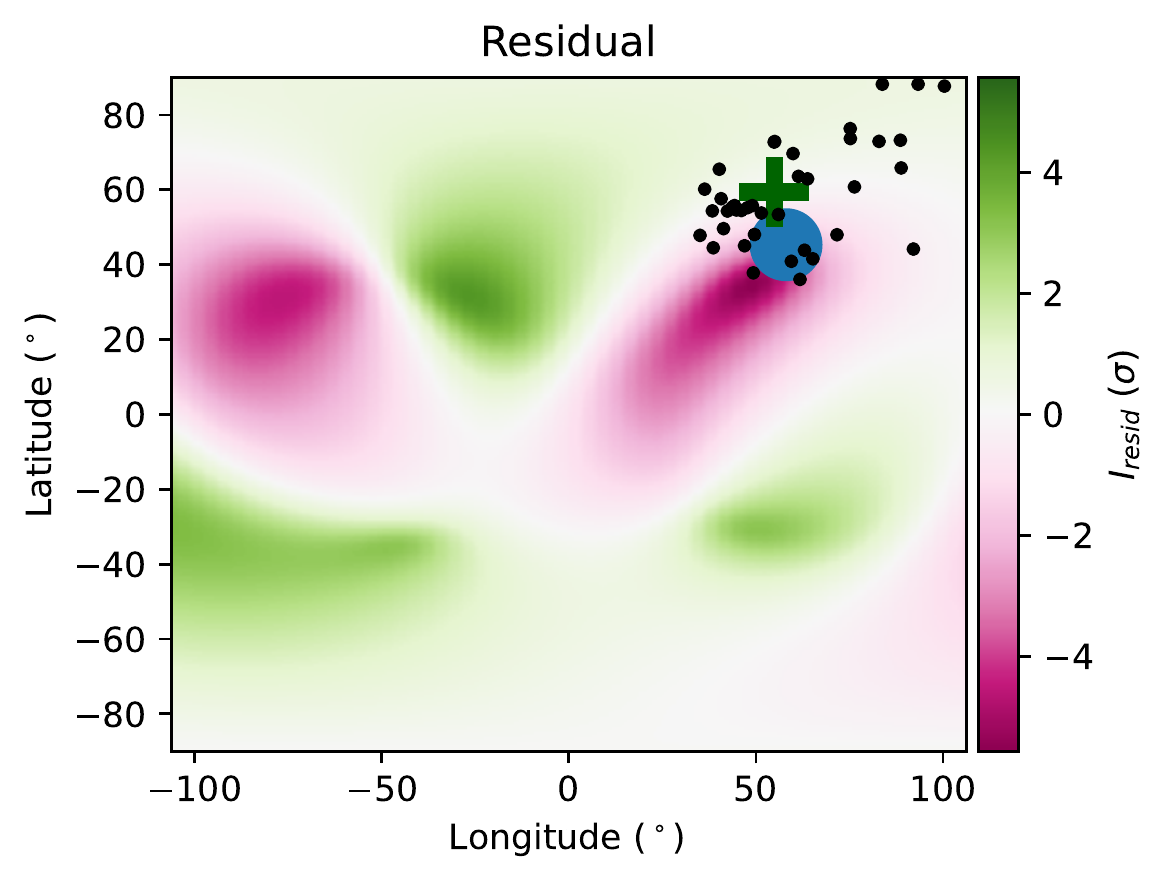}{0.30\textwidth}{Residuals with Flat Baseline}
	}
\gridline{
	\fig{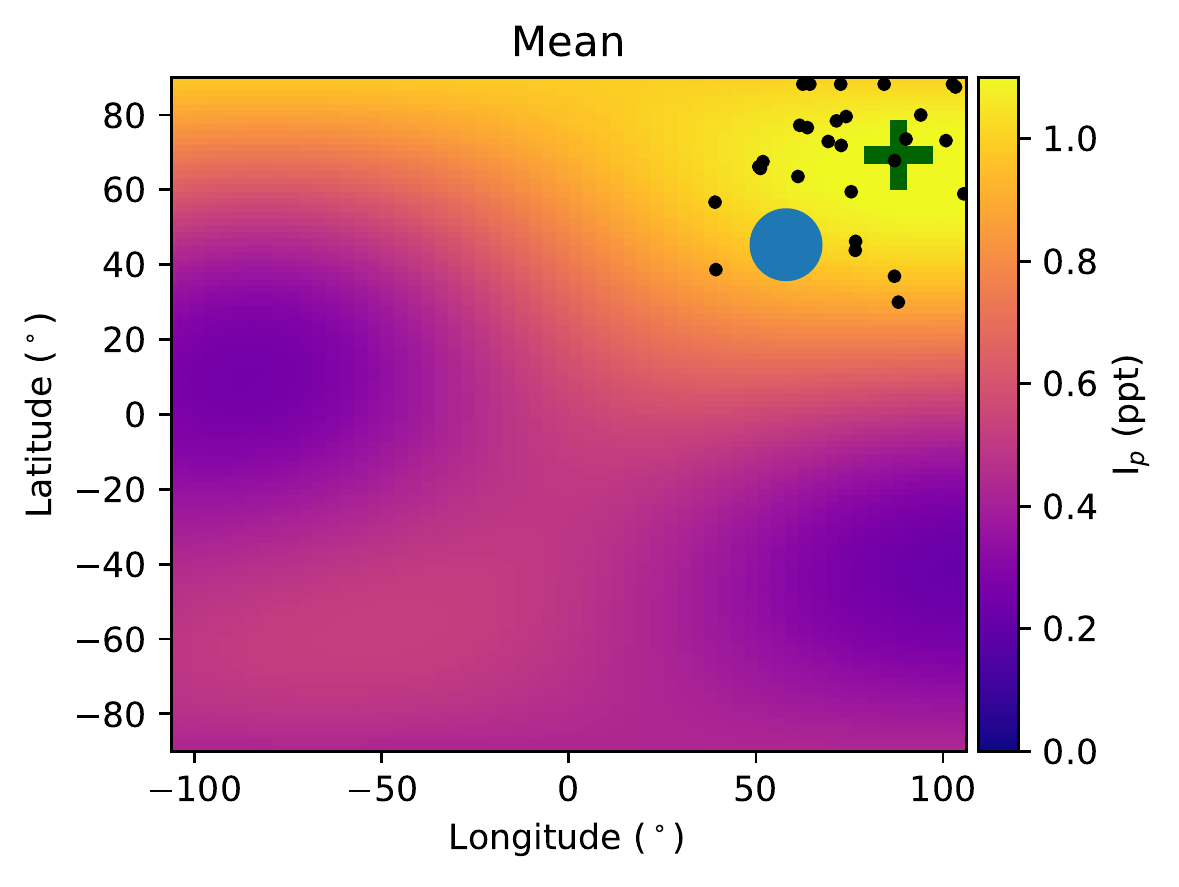}{0.30\textwidth}{Recovered \starry\ Mean with Quadratic Baseline}
	\fig{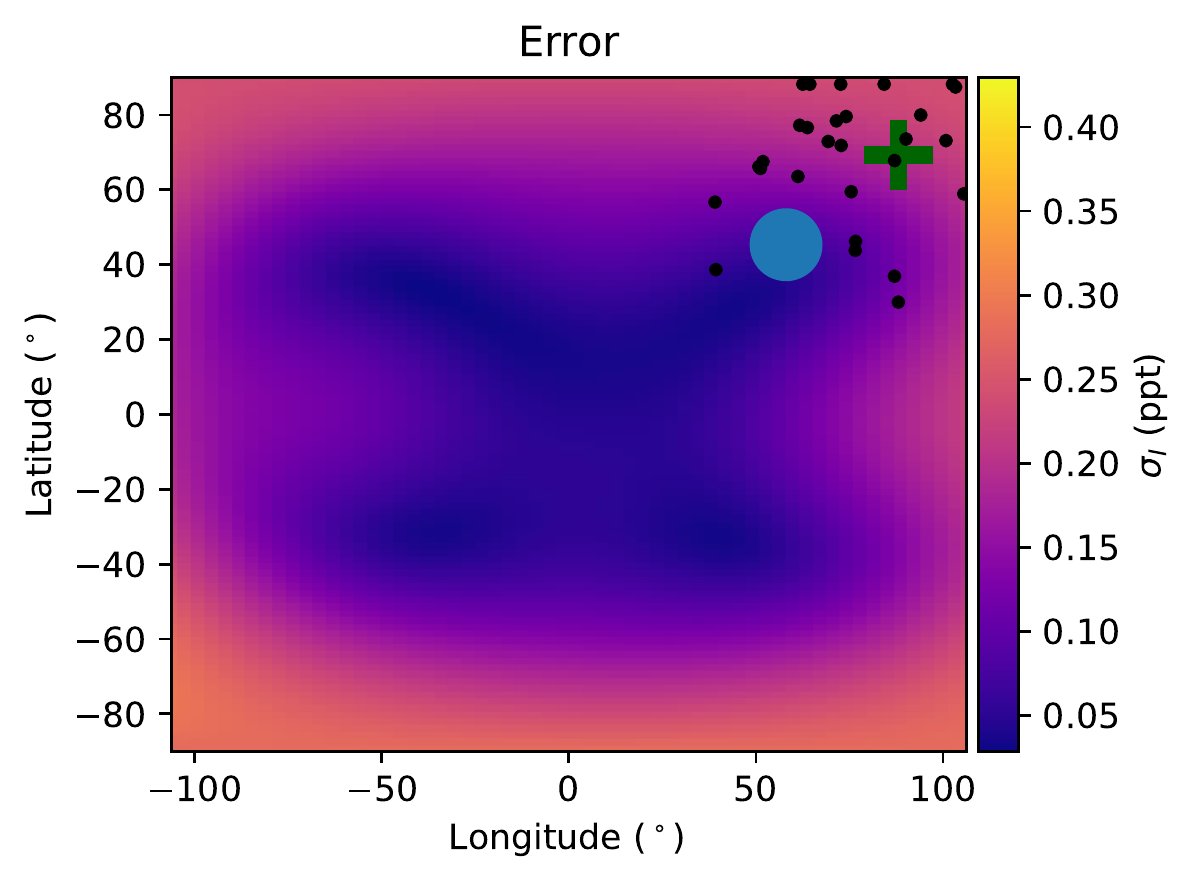}{0.30\textwidth}{Recovered \starry\ Uncertainty with Quadratic Baseline}
	\fig{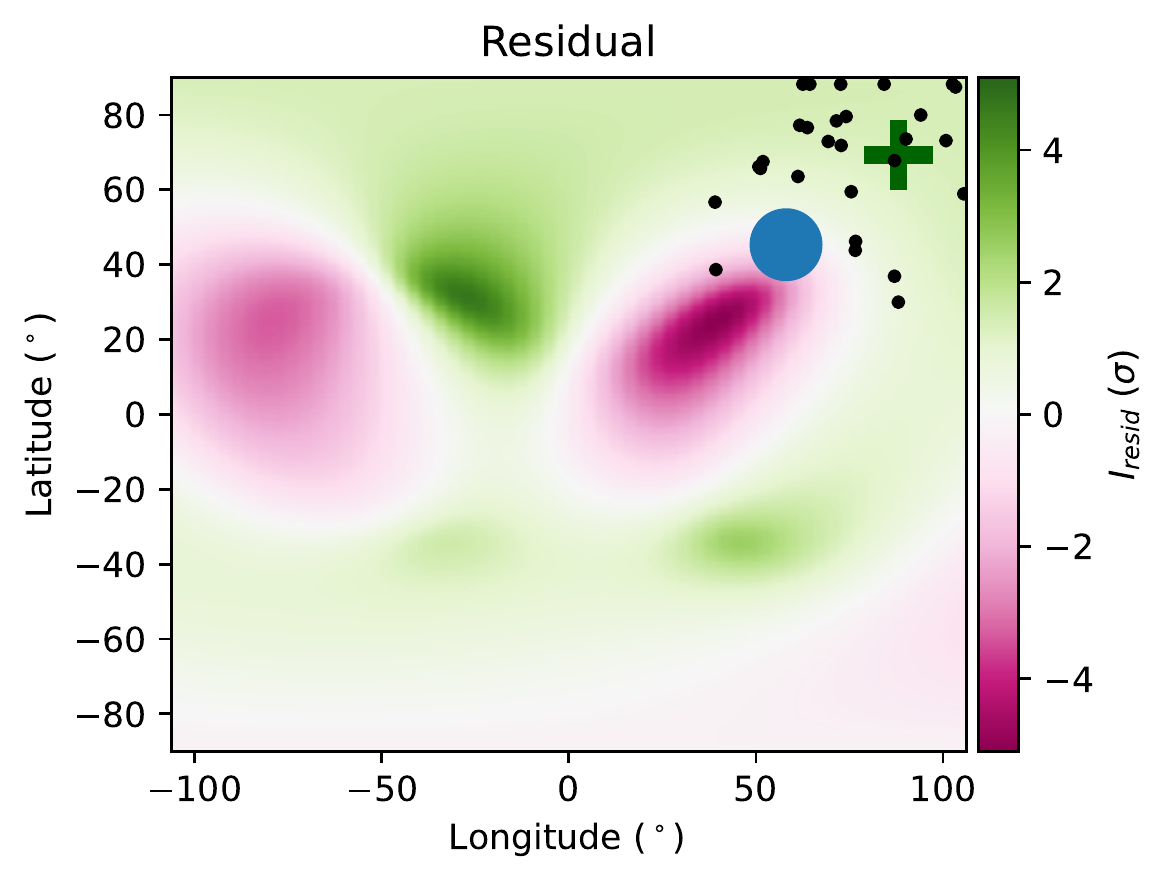}{0.30\textwidth}{Residuals with Quadratic Baseline}
	}
\caption{{\it Top:} As in Figure \ref{fig:starryMapFitsIdealized}, the arbitrary forward map is shown with the location of the peak brightness as a blue circle.
{\it Middle:} \starry\ map mean and uncertainty for the \hdb\ where the baseline is flat showing low errors (less than 25\%\ along the equator and the best constraint at the substellar point.
{\it Bottom:} \starry\ map mean and uncertainty for \hdb\ where the lightcurve has a quadratic polynomial trend, which decreases the precisions at the Eastern and Western limbs.
As in Figure \ref{fig:starryMapFitsIdealized}, we show the locations of peak brightness for individual map draws as black points and the location of peak brightness of the mean map as a plus symbol.\label{fig:starryMapFits189}}
\end{figure*}

\begin{figure*}
\gridline{	
\fig{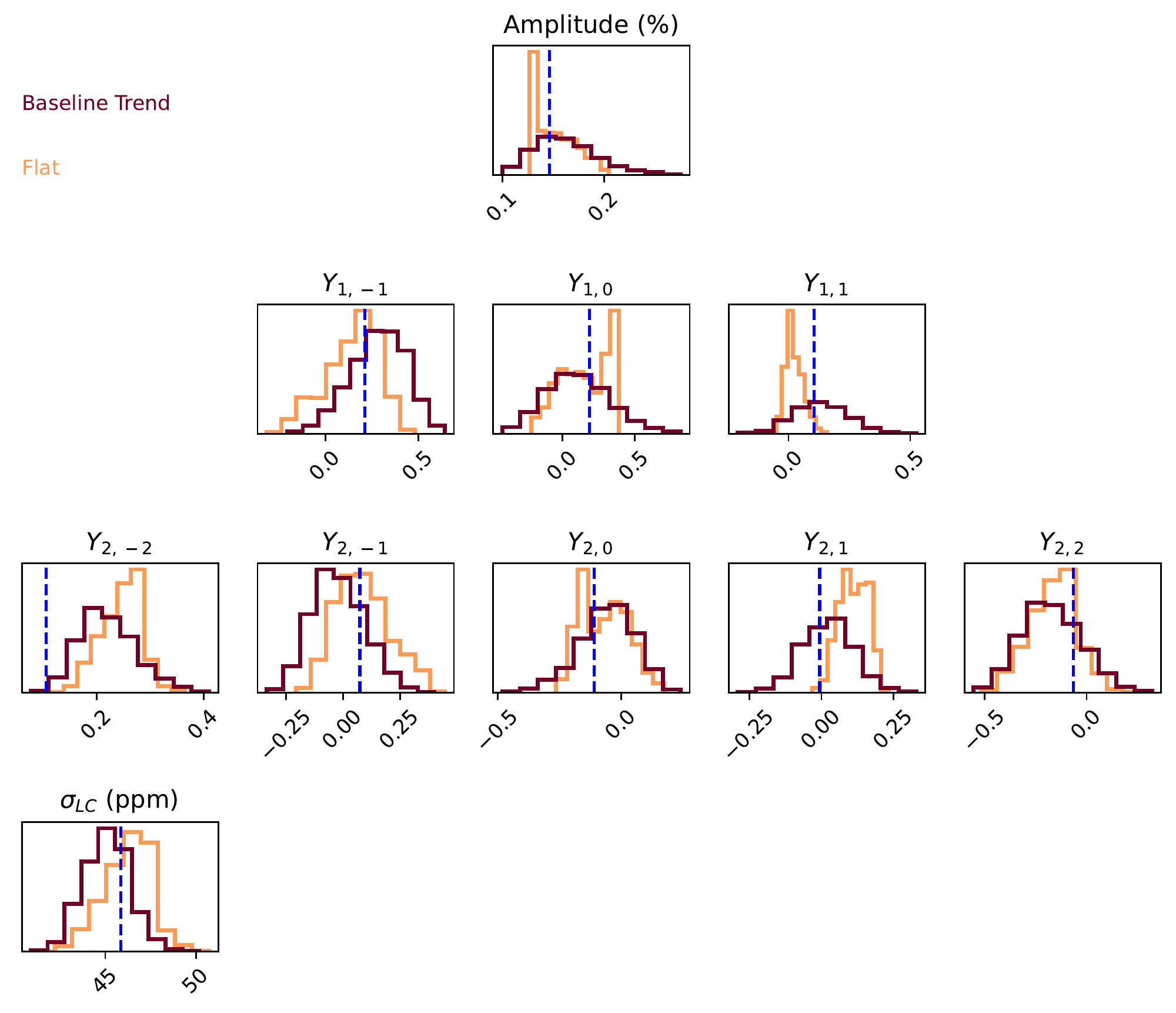}{0.99\textwidth}{Posterior Distributions for A Simulated \hdb\ Observation}
	}
\caption{Posterior distributions for all mapping variables and the lightcurve standard deviation.
The posteriors for a flat baseline are shown in light brown whereas the posteriors with a baseline trend are shown in dark brown while the inputs for the forward model are shown as blue vertical dashed lines.
At the lower signal to noise used for this JWST \hdb\ simulation compared to the ideal case shown in Figure \ref{fig:posteriorHistIdealP}, the BIC favors a second degree spherical harmonic, but the results are similar to the idealized planet.
The m=1 spherical harmonics significantly lower precision constraints when there is a baseline trend fit with a Gaussian Process model \label{fig:posteriorHistHD189}}
\end{figure*}

\subsection{Spherical Harmonic Fitting Results - \hdb\ Simulation}\label{sec:sphHarmonicHD189res}

We next examine mapping results with simulated photon and detector noise that would be present in JWST observations of a real system's brightness and orbit, but with the same arbitrary forward map as before.
We show the recovered maps for the \hdb\ simulated lightcurve and noise model in Figure \ref{fig:starryMapFits189}.
For \hdb, the map errors between a flat baseline trend and a quadratic baseline are very similar with the highest uncertainties at the poles and East and West terminators.
However, the errors are about 10\% higher with a baseline trend.

The location of peak brightness from individual map draws for the \hdb\ simulation (regardless of baseline) is also confined near the terminator of the planet and has bias closer to the North pole (here we treat North as the direction from the stellar midpoint to the closest projected location of the planet during eclipse).
This bias in peak brightness is likely related to the fact that the forward map has spherical harmonics up to third degree whereas the fit has spherical harmonics only up to second degree.
While the spherical harmonic maps are orthogonal to each other, their lightcurves are not, leading to potential biases \citep{rauscher2018moreInformativeMapping}.
We found that a lightcurve fit with third degree spherical harmonics had a closer match to the true location of peak brightness but is not justified by the Bayesian Information Criterion (BIC) due to the large number of free parameters in the fit (16 mapping terms).

We also plot the marginalized posterior histograms for all mapping variables for \hdb\ in Figure \ref{fig:posteriorHistHD189}, which can be compared to the results for the idealized planet in Figure \ref{fig:posteriorHistIdealP}.
The $Y_{1,1}$ and $Y_{2,1}$ terms are again far better constrained (in terms of the posterior width) when the baseline is known to be flat.
Interestingly, the mean values are more biased away from the truth when the baseline is known to be flat, possibly due to the 2nd order fit to the 3rd order forward model.
While the precision on the m=1 terms for the idealized planet varied by a factor of 2.8 to 5.8 between the flat and baseline trend cases, the precision for \hdb's m=1 terms only varied by 2.1 to 3.0 between the flat and baseline trend cases.
This suggests that at JWST photon and read noise precisions (31 ppm at 1 minute cadence and lower), the knowledge of the exact baseline trend is less critical than ultra-high precisions (5.6 ppm at 1 minute cadence).

One difference in the simulation of the idealized planet and \hdb\ was the polynomial inserted for the baseline.
We compared a third order polynomial (cubic) baseline into the idealized planet whereas for \hdb\ we used a second order polynomial (quadratic) baseline.
We tested whether this baseline difference affected our conclusions by using the same exact third order (cubic) polynomial for the idealized planet and \hdb.
We found that this did not affect our conclusions and the only difference was some bias in the recovered values for the map amplitude and Y$_{1,0}$ term.

\subsubsection{Fitting Baseline Trends With a Pure-Planet Model}\label{sec:baselineTrendsWithAstrophysical}
\begin{figure*}
\gridline{
	\fig{forward_map_rect_hd189733_f444w}{0.3\textwidth}{Forward Input Map}
	}

\gridline{
	\fig{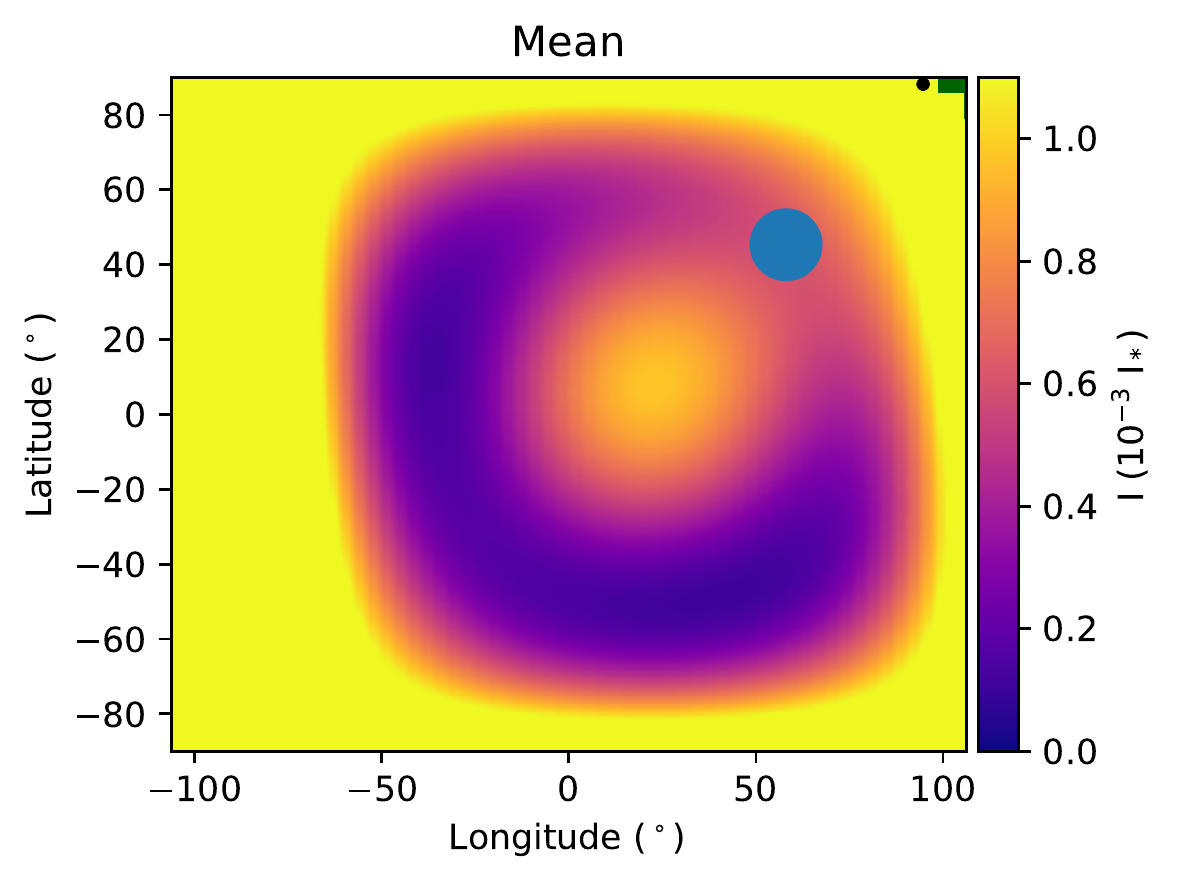}{0.30\textwidth}{Recovered \starry\ Mean with Quadratic Baseline Fit With Astrophysical Model}
	\fig{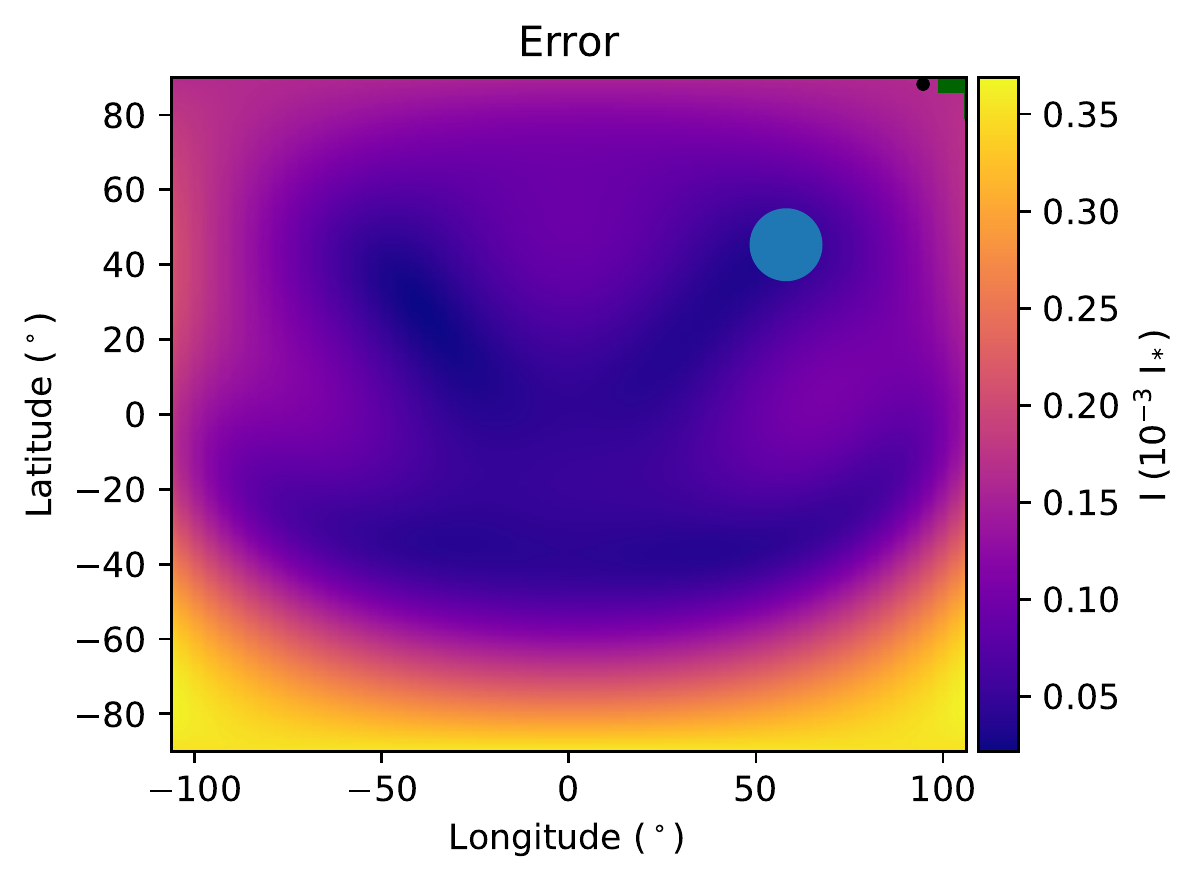}{0.30\textwidth}{Recovered \starry\ Uncertainty with Quadratic Baseline Fit With Astrophysical Model}
	\fig{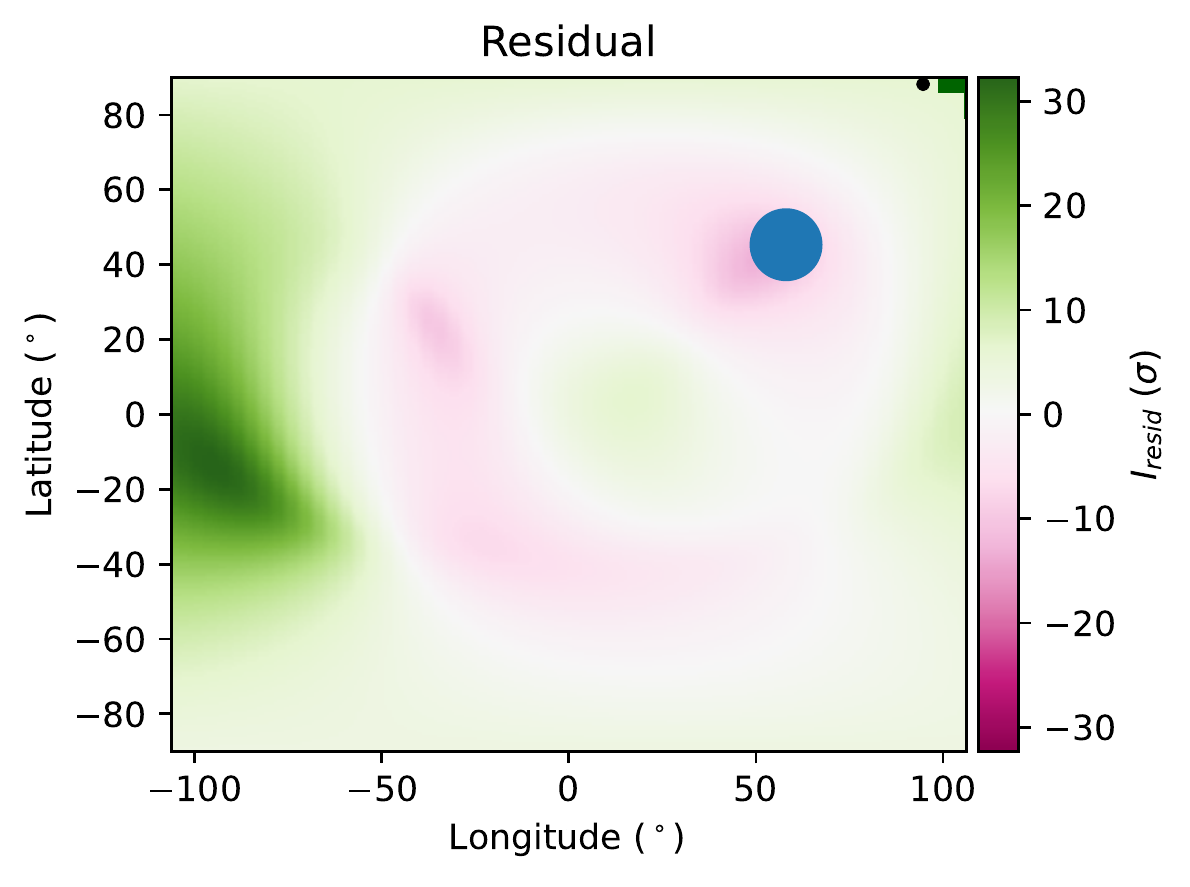}{0.30\textwidth}{Residuals with Quadratic Baseline Fit With Astrophysical Model}
	}
\gridline{
	\fig{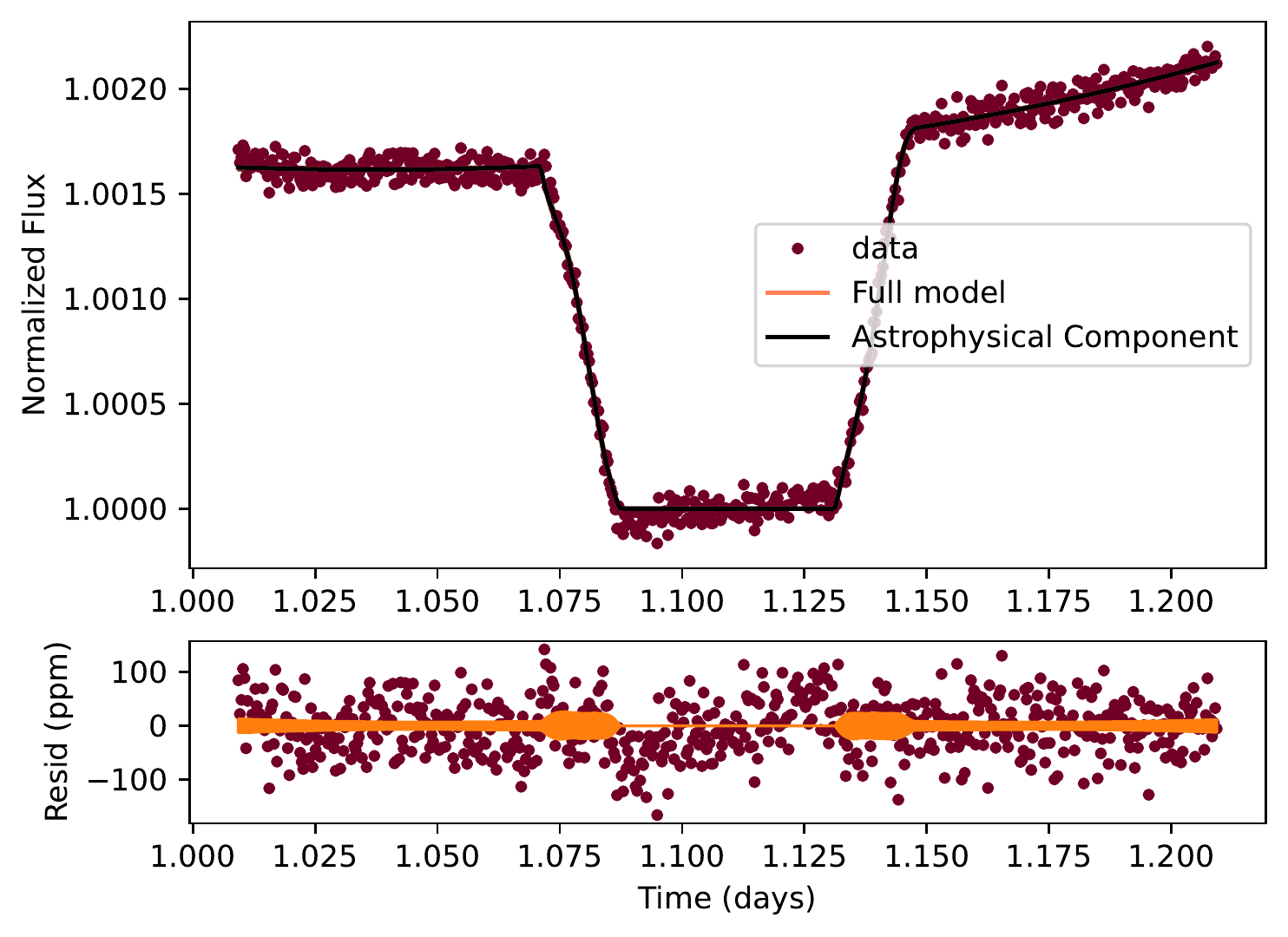}{0.5\textwidth}{Posterior Lightcurve Distribution}
	}
\caption{{\it Top:} Arbitrary forward map model for the \hdb, the same shown in Figure \ref{fig:starryMapFits189}.
{\it Middle:} \starry\ map mean, uncertainty and residuals for \hdb\ where the model has no Gaussian process or polynomial component and attempts to fit the baseline trends with planet map components only and thus has an incorrect map.
{\it Bottom:} The posterior lightcurve and residuals are most noticeable during eclipse where the lightcurve should be flat, thus indicating that a baseline model such as a Gaussian process or polynomial is favored in fitting the lightcurve, as also revealed by the two BICs of 686.1 in Figure \ref{fig:sphHarmonicLCs} (right) and 824 in in this Figure.}\label{fig:lcFitsNoGPsphericalHarmonics}
\end{figure*}

We also experimented with fitting the lightcurve with spherical harmonics but do not model any baseline trends as separate parameters and use only an astrophysical model to fit the lightcurve.
We apply this fit to \hdb\ with the injected quadratic trend.
As in Section \ref{sec:sphHarmonicHD189res}, we use second degree spherical harmonics but no Gaussian process model.
As can be seen in Figure \ref{fig:lcFitsNoGPsphericalHarmonics}, the general baseline trend can be fit with mapping features.
In particular, bright features near the terminator of the planet can produce approximate baseline trends.
However, this leaves small but significant residuals during eclipse when the planet is blocked by the star.

The lightcurve residuals of an eclipse mapping fit, especially during mid-eclipse, give a clue about whether the baseline is needed to fit the lightcurve because the planet is completely obscured for a non-grazing eclipse.
The BIC for a model with a Gaussian process fit with a lightcurve shown in Figure \ref{fig:sphHarmonicLCs} (right) and mapped in Figure \ref{fig:starryMapFits189} is 686.1.
The BIC for an astrophysical-only model that has no baseline trend parameters shown in Figure \ref{fig:lcFitsNoGPsphericalHarmonics} is 824.
Thus, a Gaussian process with a baseline trend is favored over modeling the trends from the erroneous map features.
Ignoring the residuals will give an incorrect map with larger errors and the residuals during the eclipse of the planet are important for distinguishing planet map trends from stellar or systematic trends.
Finally, another warning sign when fitting lightcurves is that the map has bright features right at the terminator of the planet.
These terminator features can approximate the stellar or systematic trends in a lightcurve so if eclipse mapping returns strong features at the terminator, caution may be warranted.
For hot Jupiters, where General Circulation Models predict a single large hotspot in infrared thermal maps \citep{showman2002circulation51peg}, terminator features are also less likely.
However, clouds can produce sharp map features near the terminator \cite{parmentier2016transitionsClouds}, so distinguishing the trends from these mapping features may be more challenging in optical wavelengths.

\subsection{Variations on Our Simulations}\label{sec:variationsOnSimulation}
So far, we have explored two different simulations of the same forward map with different precisions and two polynomials but it is natural to wonder how much our specific simulations generalize to other kinds of forward maps, lightcurve systematics and orbital parameters.
In this section, we explore some of these effects with a few examples.

\subsubsection{Comparison of Conclusions on A Different Forward Map}\label{sec:differentForwardMap}

We have found that the overall map precision is similar whether or not there is a baseline trend but that the m=1 spherical harmonics and the longitudinal precision are worse when the baseline trend is present.
We tested in Section \ref{sec:sphHarmonicHD189res} whether this conclusion is true for different baseline trends and orders (quadratic and cubic) but all on the same arbitrary forward map of Earth.
We also tested whether this conclusion depends on the specific forward map by using a GCM brightness model for \hdb\ and simulating a lightcurve with and without a baseline trend.
We then fit these two forward simulations, just like in Section \ref{sec:sphHarmonicHD189res}.
For the forward model, we use the RM-GCM developed for hot Jupiters in \citet{rauscher2010atmosphericFlows3D} and \citet{roman2019cloudsWRadiativeFeedbackHotJups} that has several improvements \citep{malsky2023polychromaticGCMs}.
One improvement is a ``picket fence'' radiative transfer scheme \citep{chandrasekhar1935radiativeEqBlanketingEffect,parmentier2014nonGreyAnalyticalModelDerivation} that treats the line and molecular opacities and the continuum opacities as separate.
Another addition to the new model is a more robust cloud parameterization that improves upon \citet{roman2021clouds3DthermalStructures}.
The forward map for the GCM simulation is shown in Figure \ref{fig:mapPosteriorsGCM} (top).

We add the same cubic polynomial baseline trend as in Section \ref{sec:forwardModel} to see how the different forward map can change the results.
The resulting precisions are very similar in that the location of peak brightness is less well constrained when there is a quadratic baseline trend.
The map at the equatorial regions near +/- 100 degrees longitude (ie. where the planet midpoint faces the observer's line-of-sight at the beginning and end of the lightcurve) are also less well constrained, as seen in Figure \ref{fig:mapPosteriorsGCM}.
This is also where the measured flux before ingress and the flux after egress is likely impacted by the baseline trend.
As with Section \ref{sec:sphHarmonicHD189res} the overall uncertainty map looks similar whether or not there is a baseline trend.
We do note that the precision of the peak brightness location is more adversely affected by the cubic baseline with the GCM forward map than the arbitrary Earth map.
In our arbitrary forward map, the location of peak brightness is off to the corner of the map, which is recovered whether or not there is a baseline trend.
However, with the GCM forward map, the difference in precision is more noticeable.
We also note that some map draws from the posterior distribution have their locations of peak brightness at the extremes of our priors at $\pm$ 90 degrees latitude.
There are more of these outlier map draws in the posterior when there is a baseline trend.

The posterior distributions for the spherical harmonics are shown in Figure \ref{fig:posteriorHistHD189GCM}, along with the true input values from the GCM forward map.
As with Figure \ref{fig:posteriorHistHD189}, the precision on the m=1 spherical harmonics is significantly better when the baseline is known to be flat.
However, the other spherical harmonic terms have similar posterior distributions, as can be seen when comparing Figures \ref{fig:posteriorHistHD189} and \ref{fig:posteriorHistHD189GCM}.

\subsubsection{Uncertain Orbital Parameters}\label{sec:differentForwardMap}
\citet{deWit2012eclipsemap189} find that uncertain orbital parameters can impact the resulting brightness maps for \hdb\ and that there is a correlation between the eccentricity, impact parameter, stellar density and the brightness distribution.
The uncertainty can be reduced by radial velocity measurements for a well-characterized planet like \hdb.
Furthermore, \citet{rauscher2018moreInformativeMapping} find that the orbital parameter uncertainties have a small effect on the first five orthogonal mapping components.
The coefficients on the mapping components are changed by $\lesssim$1\% by orbital uncertainties versus fixed uncertainties for \hdb.
Uncertain orbital parameters (other than eccentricity) also have a negligible effect on the eclipse map recovered for WASP-18 b, with maps well within uncertainties even when the impact parameter, semi-major axis and eclipse time were varied by $\pm 1\sigma$ \citep{coulombe2023wasp18EmissionSpec}.

We tested a fit for \hdb\ where the inclination and stellar mass were varied as additional variables using an inclination uncertainty from \citet{agol2010hd189} and the stellar mass uncertainty from \citet{addison2019minervaAustrialisFirstResults}.
We find that the overall map error and mapping term posteriors widths were very similar whether the orbit was held fixed or had a variable inclination and stellar mass, but with a larger bias in the posterior for the Y$_{1,0}$ and Y$_{2,-2}$ terms.
However, our conclusion that a non-flat baseline mostly affected the m=1 terms (Y$_{1,1}$ and Y$_{2,1}$) is unchanged for the variable orbit scenario.
\hdb, with its long history of high precision characterization from radial velocity and photometric monitoring may be less affected than planets with weaker constraints on the orbit.

\subsubsection{Impact Parameter Choices}\label{sec:impactParamChoice}
\begin{figure}
    \centering
    \gridline{
    \fig{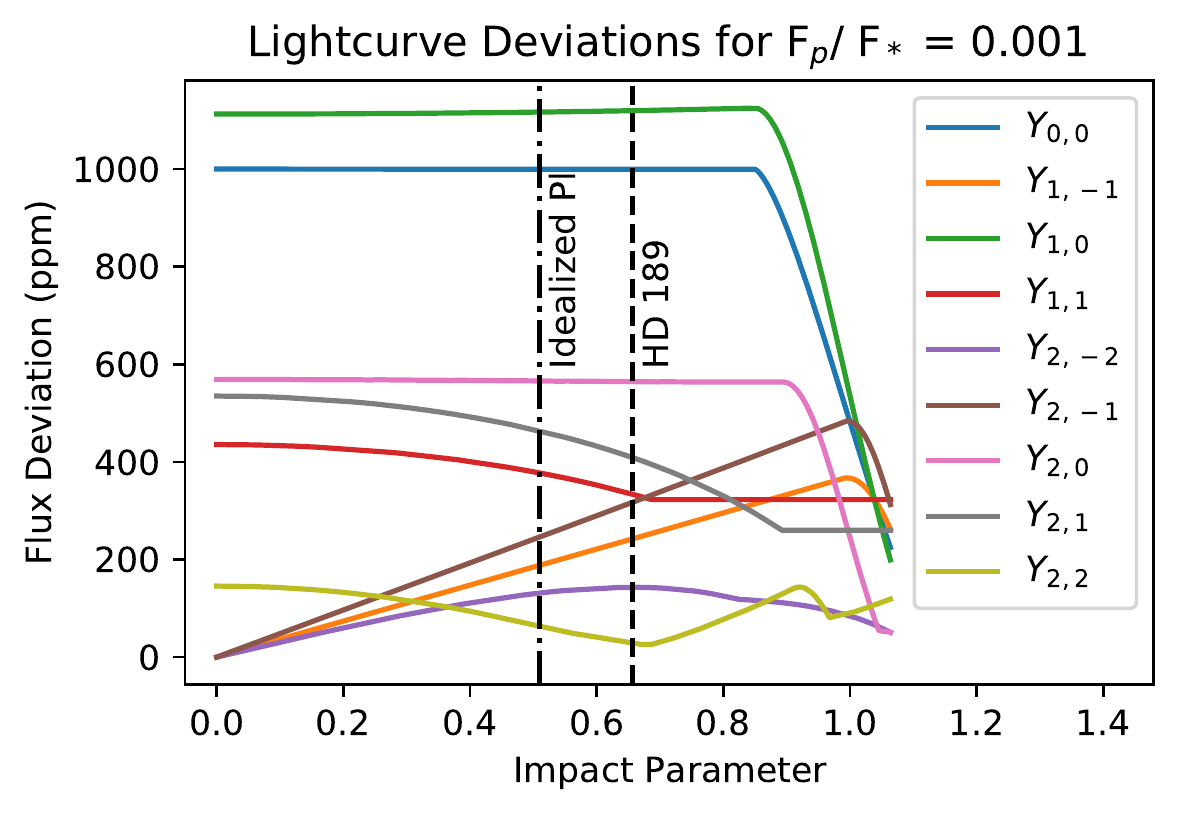}{0.49\textwidth}{}
    }
    \caption{
    Each curve shows the variation during the eclipse lightcurve due to one spherical harmonic component if the integrated flux of the planet with that spherical harmonic component was 0.001 times the integrated flux of the star.
    We selected impact parameters with b$>$ 0.5 in order to have significant contributions from the latitudinal terms Y$_{1,-1}$, Y$_{2,-1}$ and Y$_{2,-2}$.}
    \label{fig:impactParamDeviation}
\end{figure}

Our examples were for similar impact parameters of 0.51 (idealized planet) and 0.66 (\hdb).
As shown in Figure \ref{fig:impactParamDeviation}, the relative contributions each spherical harmonic depend on the impact parameter, $b$.
When $b \lesssim 0.1$, the latitudinal components (Y$_{1,-1}$, Y$_{2,-2}$ and Y$_{2,-1}$) contribute either negligibly (or zero) to the lightcurve so the latitudinal offsets and gradients cannot be constrained from the lightcurve.
When $b=0$, these latitude components are in the ``null space'' \citep{cowan2013lcInclinationObliquityAlbedo,luger2021mappingStellarSurfacesDegeneracies} and can be any value while still resulting in mathematically identical lightcurves (though some constraints on them can be enforced by ensuring a nonnegative planet map).

We briefly test the sensitivity of our conclusions to the impact parameter by calculating a forward model that has zero impact parameter.
We summarize the results in this subsection but do not plot all posteriors.
We simulated a lightcurve that has the same systematics and parameters as for \hdb\ described in Section \ref{sec:forwardModel} but with $b$=0.
We find that the location of peak brightness is poorly constrained in the maps with a large range of latitudes below and above the equator.
However, whether the baseline is assumed to be flat or has the quadratic baseline added, the posteriors on the first and second degree spherical harmonics overlap with the exception for the m=1 terms Y$_{1,1}$ and Y$_{2,2}$.
In other words, our conclusions for the zero-impact parameter planet model are the same as for \hdb\ in Section \ref{sec:sphHarmonicHD189res}.
Whether the impact parameter for a \hdb-like planet was assumed to be $b=0.66$ or $b=0.0$, the ratio of uncertainty on the m=1 terms drops by a factor of $\sim$3 for the Y$_{1,1}$ term and $\sim$2 for the Y$_{2,1}$ term.

\subsubsection{Baseline Choices}\label{sec:baseChoices}
Another consideration we have not discussed is how our results depend on the baseline polynomial order beyond 3 or with different magnitudes of baselines.
We do not perform an exhaustive simulation of all possible baselines, but use a GP forward model as an illustrative example.
For the Gaussian process forward model we chose a SHOT kernel function in \texttt{celerite2} with a highly dampened oscillator that has a quality factor (Q) of 0.25, a timescale ($2\pi/\omega_0$) of 17 hours, where $\omega_0$ is the undamped angular frequency, and a standard deviation of 1000 ppm.
We recovered the maps as described in Section \ref{sec:sphHarmonicHD189res}.

In our test with a GP forward model, we found again that the m=1 spherical harmonic modes were most affected by a non-flat baseline.
The uncertainty on these coefficients was 4.1 and 2.8 times larger for $Y_{1,1}$ and $Y_{2,1}$ respectively than when the baseline is known to be flat.
This is larger than the factor of 3.0 and 2.1 for the same coefficients when using a quadratic baseline described in Section \ref{sec:forwardModel}.
Additionally, the GP forward model had larger uncertainties on latitudinal components, with uncertainties on $Y_{1,-1}$ that were 1.7 times larger than the flat case as compared to 1.2 times larger with the quadratic baseline forward model.
The larger uncertainties for additional terms are likely caused by high frequency noise that is not present in the quadratic baseline forward model.
The results of this paper apply mainly to long-term baseline trends as compared to the ingress and egress timescale (24 minutes for \hdb).
Additional analysis of correlated noise on a timescale less than the ingress and egress time can destroy more information on derived eclipse maps.

\subsection{Eigenmap Fits for \hdb}\label{sec:eignmapFits}

We also use the eigencurve method to fit the map of \hdb\ using mapping components with orthogonal (and thus uncorrelated) lightcurves \citep{rauscher2018moreInformativeMapping}.
We use the Three-dimensional Exoplanet Retrieval from Eclipse Spectroscopy of Atmospheres (ThERESA) code designed for multiwavelength mapping \citep{challener2021ThERESA} but use it with a single wavelength here on the simulated data for \hdb. 
The 2D mapping with ThERESA uses a similar framework as \citet{rauscher2018moreInformativeMapping}, but uses truncated singular-value decomposition (TSVD) to generate eigencurves.
TSVD has the advantage of not doing a mean-subtraction of the lightcurve which was required in the principal components analysis (PCA) approach in \citet{rauscher2018moreInformativeMapping}.
This eigencurve method is different from Sections \ref{sec:sphHarmonicFits} through \ref{sec:sphHarmonicHD189res}, where the lightcurves from the spherical harmonic components have significant correlations, as shown in Figure \ref{fig:posteriorCorner}.
However, in both the ThERESA and spherical harmonic fit approaches, we enforce a positive flux constraint at the visible longitudes.

\begin{figure}
    \centering
    \includegraphics{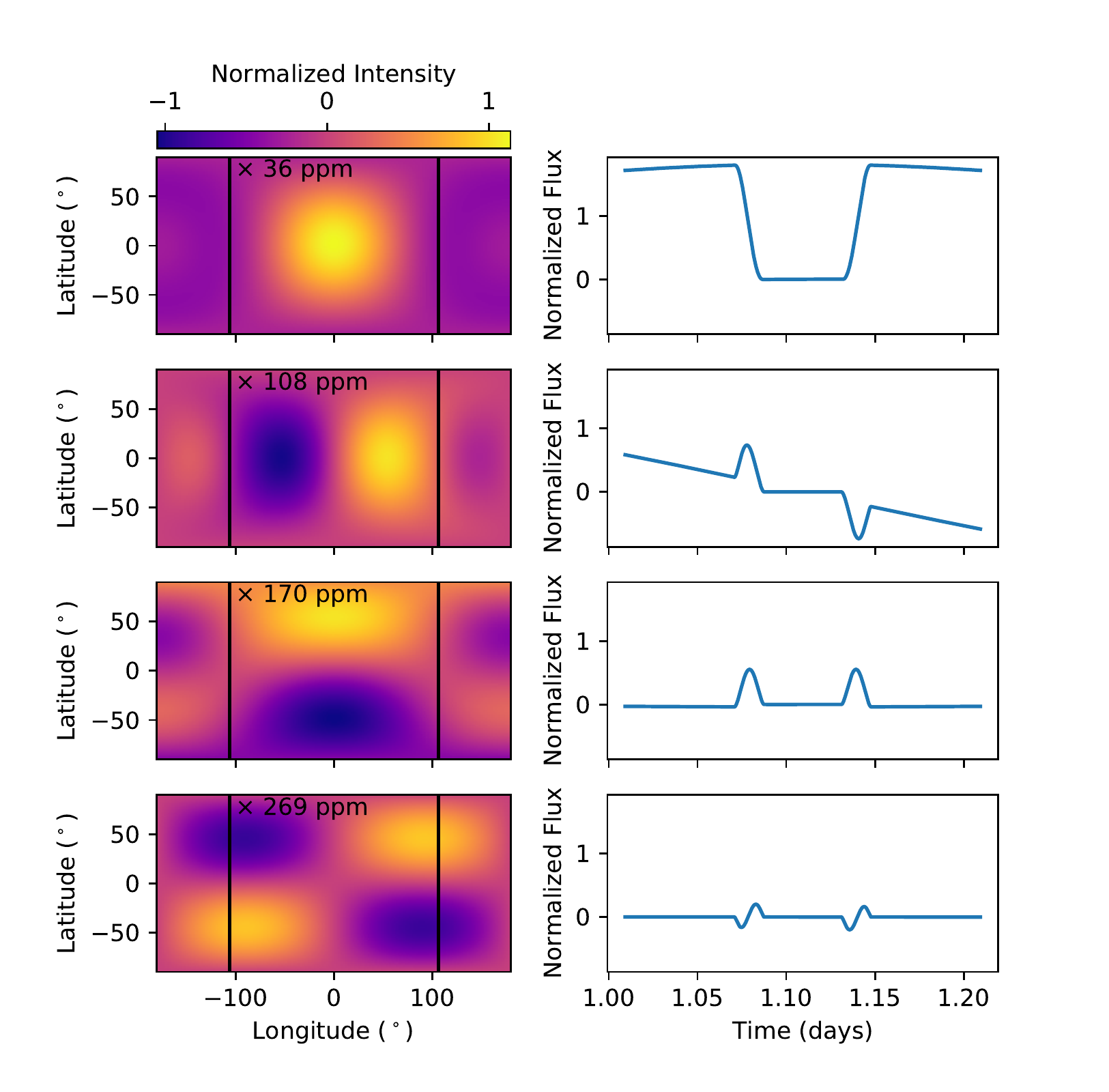}
    \caption{Eigenmap components (left) of the best-fitting ThERESA model when using a quadratic baseline and their associated eigencurves (right). The eigenmaps are normalized to highlight differences in structure, with the best-fitting coefficients listed with the corresponding eigenmaps. The vertical black lines denote the range of longitudes visible during the observation.}
    \label{fig:ThERESAEigenmaps}
\end{figure}

ThERESA optimizes the maximum spherical harmonic complexity $l_{\rm max}$ and the number of eigencurves $N$ by minimizing the Bayesian Information Criterion.
We tested all valid combinations of $l_{\rm max} \leq 5$ and $N \leq 6$, and achieved the best fit with $l_{\rm max} = 2$ and $N = 4$. 
Counting the uniform ($Y_{0,0}$) map term, a stellar correction term, and two quadratic baseline terms, the model has eight free parameters (and fewer with less complex baseline models discussed below).
The $N=4$ eigenmapping terms and their associated lightcurves are shown in Figure \ref{fig:ThERESAEigenmaps}.
The orthogonal nature of the lightcurves results means that there are fewer parameters to fit than the spherical harmonics and that their coefficients are not correlated.
The eigenmaps are also ordered in terms of the singular values with the largest structures (first eigenmap) having the largest change in the lightcurve and the smallest structures (fourth eigenmap) having the smallest changes in the lightcurve, as shown in Figure \ref{fig:ThERESAEigenmaps}.

\begin{figure}
    \centering
    \includegraphics{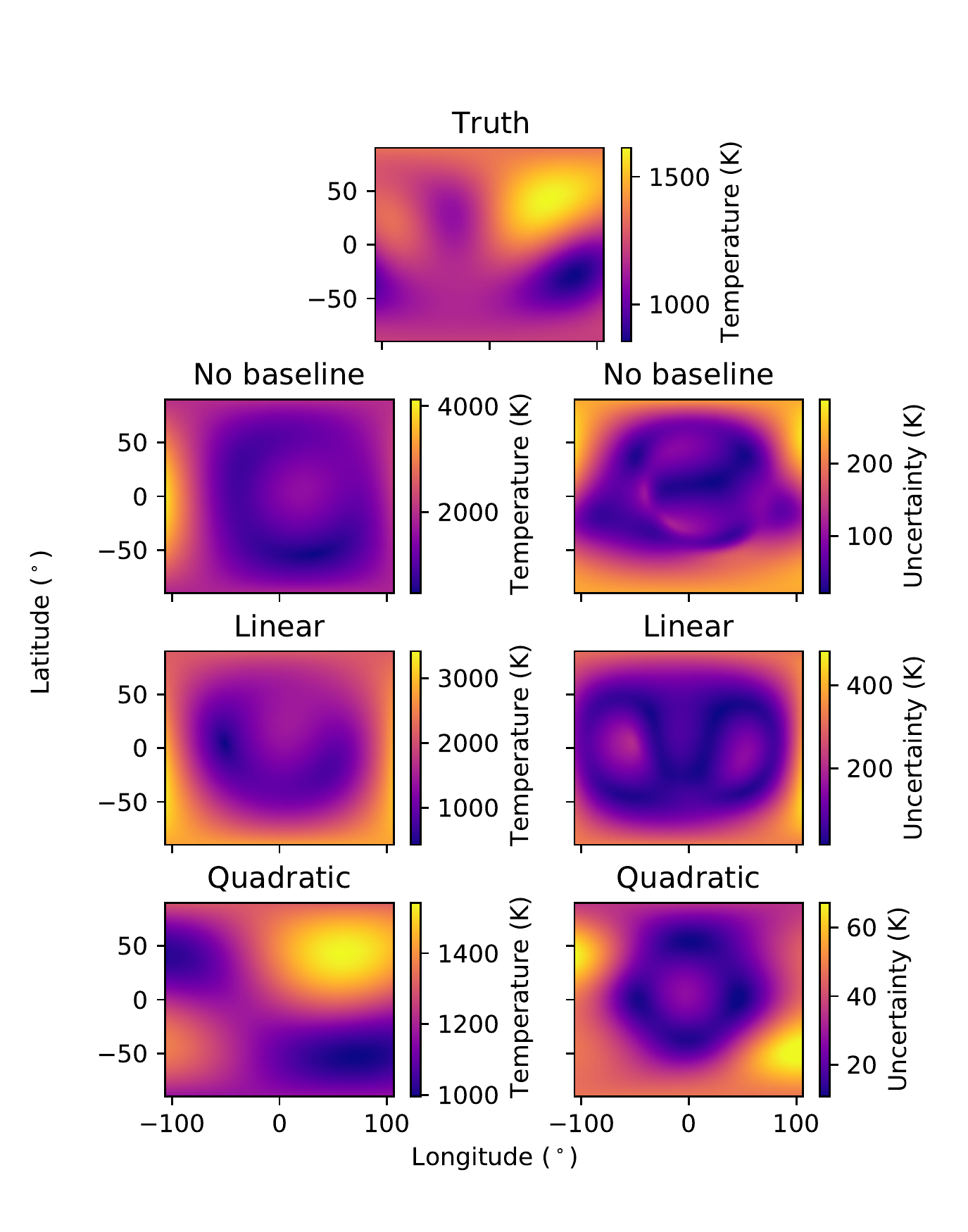}
    \caption{Best-fitting maps using ThERESA \citep{challener2021ThERESA} with different baseline models. We have restricted the $x$ axis to only show longitudes visible during the observation (-106.2 -- 106.2$^\circ$). Note the different scales on the colorbars.\label{fig:ThERESABestFit}} 
\end{figure}

For the ThERESA lightcurve fitting, we experimented with flat, linear and quadratic polynomial baselines to model the systematic trends.
This is different from the Gaussian process model used in Sections \ref{sec:sphHarmonicFits} through \ref{sec:sphHarmonicHD189res}.
The resulting map fits are visible in Figure \ref{fig:ThERESABestFit} for the three assumptions about the systematic trends.
We first show the assumption that there is no baseline and that we get {\it very} incorrect maps.
The same holds true with a very incorrect map for an assumption that the baseline trend is linear.
This is because the best fit model re-creates lightcurve trends and curvature by including large ($>$3000 K brightness temperature) peaks near the planet's terminator that reproduce the baseline trends.
In other words, the stellar or instrument trends can be incorrectly modeled with planet map features as in Figure \ref{fig:lcFitsNoGPsphericalHarmonics}.
Baseline trends also may appear to have the same shape as the phase curve observations such as with the Spitzer Space Telescope \citep[e.g.][]{knutson2007map189}.
In the case of HAT-P-7 b, baseline trends due to stellar supergranulation can look just like phase curve variations.
Thus, this planet variability has been interpreted as either weather patterns \citep{armstrong2016hatp7variability} or stellar supergranulation \citep{lally2022hatp7reassessingVariability}.

In contrast to the flat or linear assumption, a quadratic baseline trend correctly recovers the input truth map up to the resolution allowed by the 4 mapping terms.
At the same time, the model with a quadratic baseline is favored by the BIC (676.4) over the no baseline (807.4) and linear baseline (696.7) fits.
The recovered maps contains a similar peak brightness location as the forward map (59.6$\degree$, 48.0$\degree$).
The map maximum location is constrained to be ($59.4^{+7.6}_{-6.9} {\degree}$, 42.5$^{+1.6}_{-1.4}{\degree}$) from the ThERESA fit.
Thus, the longitude is accurate to within error and the latitude has a 3.5$\sigma$ offset.
The ThERESA map also accurately depicts the relatively dark regions near (-30$\degree$,30$\degree$) and (80$\degree$, -30$\degree$) in the forward map with some offsets in location.

We note that the maximum visible longitudes are -106.2$\degree$ to 106.2$\degree$ so we only plot this region of the covered maps.
Outside of these longitudes, the map is unconstrained and there is no expectation that the recovered map should agree with the forward map.
ThERESA does not recover all the features of the forward map because the fit is restricted, by the data quality, to only four eigenmap components (Figure \ref{fig:ThERESAEigenmaps}), and, thus, can only constrain large-scale features.
For example, the fourth eigenmap allows for the northeastern brightness peak and colder southeastern quadrant, but simultaneously creates the inverse in the western hemisphere, a mismatch with the forward map.
With lower observational uncertainties, we could be justified in fitting to smaller-scale features.

Comparing Figure \ref{fig:ThERESABestFit} (bottom row) and Figure \ref{fig:starryMapFits189} (bottom row) shows that both the spherical harmonic \starry\ fit and the eigencurve ThERESA fit methods recover the same qualitative map structure and map uncertainties.
Note that near the forward model peak at (59.6$\degree$, 48.0$\degree$), the ThERESA brightness temperature uncertainty of 15 K at 1450 K is an 8\% intensity uncertainty at 4.4 $\mu$m, which is similar to the 8\% intensity uncertainty with the \starry\ spherical harmonics.
With both methods, the location of peak brightness is in the correct quadrant with respect to the inclination of the planet orbital plane.\footnote{There is an ambiguity about the North or South directions when eclipse mapping because we cannot determine whether the star's midpoint is eclipsing the Northern or Southern hemisphere of the planet.
In this paper, we assume that the planet's North Pole up and that the planet transits below (south of) the stellar mid point and eclipses above (north of) the stellar midpoint.}
They also both have errors that grow toward the limbs of the planet.
However, the ThERESA modeling approach shows a small peak of uncertainty at the substellar point (0$\degree$,0$\degree$) surrounded by a low noise ``doughnut" about 40$\degree$ in radius.

	

\section{Conclusions\label{sec:conclusions}}
Mapping planets requires information about both the eclipse events as well as the overall flux variations from planet rotation.
If there are baseline trends in the lightcurve that are unrelated to the planet, such as stellar spot rotation or detector systematics, some of the planet information becomes correlated or lost.
This paper assessed how long-term baseline trends can affect eclipse mapping of hot Jupiter planets.
We created simulated lightcurves for an idealized planet with a precision of 5.6 ppm per minute as well as a \hdb\ planet model as simulated for the JWST NIRCam instrument with a precision of 31 ppm per minute.
We fit these lightcurves with 2nd and 3rd degree spherical harmonic maps respectively with a condition that the map is nonnegative at visible longitudes.
We then added long-timecale ($\gtrsim 1 hr$) systematic trends to the lightcurves to learn how the trends affected the map precisions.

Modeling the lightcurve with pure astrophysical map variations can give very wrong maps because the baseline trend will be erroneously modeled with planet map features, especially at the terminator.
Fortunately, the baseline trend can be revealed during eclipse where the planet is completely obscured for non-grazing geometries.
The lightcurve with these pure-astrophysical model has larger residuals especially during the planet's mid-eclipse.
The BIC for models that include a baseline trend either with a Gaussian process or polynomial are favored over the purely astrophysical one.
Another clue that a pure-astrophysical model is wrong in the presence of baseline trends is that it has strong features near the terminator. 
It is instead necessary to fit the lightcurves simultaneously with astrophysical map features and a systematic trend model that is either a Gaussian process or polynomial.

For spherical harmonic fits, we find that the recovered m=1 spherical harmonics are most strongly impacted by the baseline uncertainty.
The impact on the m=1 map precision is stronger for the idealized planet than for \hdb\ at JWST precisions.
These m=1 spherical harmonics describe the longitudinal structure of the map and the Y$_{1,1}$ and Y$_{2,1}$ terms have significant linear trends in their baselines due to the rotation of the planet.
The Y$_{3,1}$ is flat out of eclipse but correlates significantly with Y$_{1,1}$ and Y$_{2,1}$ so they all have reduced precision if a significant baseline is present.
For the idealized planet at 5.6 ppm per minute, the difference in precision of the m=1 terms ranges by a factor of 2.8 to 5.8 if the a baseline trend is present.
However, at JWST precisions of $\gtrsim$31 ppm per minute, the m=1 spherical harmonic terms are affected at a smaller level and vary by a factor of 2.1 to 3.0 in precision due to a baseline trend.

We find that when our model allows for either a Gaussian process or polynomial trend, the mean maps recover the overall structure of the input forward model as well as the location of peak brightness.
The uncertainty map, as measured by the standard deviation of the posterior map evaluations, is also very similar whether or not there is a baseline trend in the forward model.
Thus, the change in precision for the m=1 map components gets largely washed out by uncertainty in the other spherical harmonic terms.
We find that the location of peak brightness is slightly more precise for a known flat baseline than with a baseline trend and Gaussian process fit, but it still contains a significantly broad tail of probabilities if the map priors have no constraints, such as would come from general circulation models.

We also tested modeling the \hdb\ data with ThERESA, which uses eigencurves and eigenmaps to avoid correlated parameters and extract the maximum amount of statistically-justifiable mapping information from an observation \citep{challener2021ThERESA,rauscher2018moreInformativeMapping}. 
A ThERESA model with a quadratic baseline correctly recovered the large-scale features of the forward model, using four eigencurves/eigenmaps from spherical harmonics up to second order for a total of eight free parameters, compared to 9 mapping parameters and 3 Gaussian process parameters for a total of 12 for the spherical harmonic fits for \hdb.
We find that the eclipse (vs.\ only phase-curve variation) is able to resolve any potential degeneracies between the baseline polynomial terms and the eigencurve/eigenmap coefficients, leaving parameters mostly uncorrelated.
The presence of the baseline only slightly increases the measured uncertainties on the temperature map.

These results suggest that baseline trends can impact the precision of eclipse maps at ultra high precision of far-future observatories ($\lesssim$6 ppm per minute) but less so at JWST precision on the best eclipse mapping target, \hdb.
It is important to model the baseline trends rather than fit them with erroneous map features, as guided by the residuals during eclipse.
Thus, observers planning eclipse mapping observations may strategize on where to focus telescope time.
When the m=1 terms and the longitudinal precision of the brightness peak are very important to the science case, more time should be devoted to expanding the baseline before and after eclipse and to understand the baseline shape and curvature.
When the overall map precision including the latitudinal dependence is critical to the science case, improved precision at ingress and egress (such as by observing more eclipses with short baselines) should be where telescope time is focused.
Overall, the prospects are good for recovering accurate maps of exoplanets and describing their overall structure when accounting for long timescale baseline trends with a polynomial or Gaussian process regression.


\begin{acknowledgements}

MCMC fitting makes use of \texttt{emcee} \citep{foreman-mackey2013emcee} and the covariance plot was made with \texttt{corner.py} \citep{foremanCorner}.
Funding for E Schlawin is provided by NASA Goddard Spaceflight Center.
R Challener is supported by a grant from the Research Corporation for Scientific Advancement, through their Cottrell Scholar Award. M Mansfield acknowledges support by NASA through the NASA Hubble Fellowship grant HST-HF2-51485.001-A awarded by the Space Telescope Science Institute.
Thank you to Isaac Malsky for providing a GCM simulation of \hdb.
We thank the University of Michigan Institute for Research in Astrophysics for hosting the ``Multi-Dimensional Characterization of Distant Worlds'' workshop in Ann Arbor during October 2018, where this project idea originated and this team was put together.
This research has made use of NASA's Astrophysics Data System Bibliographic Services.
We respectfully acknowledge the University of Arizona is on the land and territories of Indigenous peoples. 
Today, Arizona is home to 22 federally recognized tribes, with Tucson being home to the O'odham and the Yaqui.
Committed to diversity and inclusion, the University strives to build sustainable relationships with sovereign Native Nations and Indigenous communities through education offerings, partnerships, and community service

%

\vspace{5mm}


\software{astropy \citep{astropy2013}, 
          \texttt{matplotlib} \citep{Hunter2007matplotlib},
          \texttt{numpy} \citep{vanderWalt2011numpy},
          \texttt{scipy} \citep{virtanen2020scipy},
          \starry\ \citep{luger2019starry},
          \texttt{pymc3} \citep{salvatier2015probProgramming},
          \texttt{celerite2} \citep{foreman-mackey2018celerite},
          ThERESA \citep{challener2021ThERESA}
           }
\end{acknowledgements}



\appendix

\section{Forward model parameters}

Table \ref{tab:forwardParams} lists the parameters of the forward model.
\begin{deluxetable}{ccc}[hbt!]
\tablecaption{Summary of forward model parameters\label{tab:forwardParams}}
\tablecolumns{3}
\tablewidth{0pt}
\tablehead{
\colhead{Parameter} &
\colhead{Idealized Planet} &
\colhead{\hdb} \\
}
\startdata
$R_p$ & 0.1 R$_\odot$ & 0.115 R$_\odot$\tablenotemark{a} \\
$R_*$ & 1.0 R$_\odot$ & 0.765 R$_\odot$\tablenotemark{a} \\
b	& 0.5129	& 0.656\tablenotemark{b} \\
P	& 1.000 d	& 2.218577 d\tablenotemark{b} \\
a	& 0.01957 AU & 0.03106 AU\tablenotemark{b} \\
a/R$_*$ & 4.2083	& 8.72994\tablenotemark{b} \\
F$_p$/F$_*$ & 0.100\%	& 0.14619\% \\
$Y_{0,0}$  & 1.0\tablenotemark{a} & 1.0\tablenotemark{a} \\
$Y_{1,-1}$ & 0.2126  & 0.2126   \\
$Y_{1,0}$  & 0.1876  & 0.1876   \\
$Y_{1,1}$  & 0.1042  & 0.1042   \\
$Y_{2,-2}$ & 0.1049  & 0.1049   \\
$Y_{2,-1}$ & 0.074   & 0.074    \\
$Y_{2,0}$  & -0.1099 & -0.1099  \\
$Y_{2,1}$  & -0.0062 & -0.0062  \\
$Y_{2,2}$  & -0.0655 & -0.0655  \\
$Y_{3,-3}$ & 0.1211  & 0.1211   \\
$Y_{3,-2}$ & 0.1538  & 0.1538   \\
$Y_{3,-1}$ & -0.0459 & -0.0459  \\
$Y_{3,0}$  & 0.0555  & 0.0555   \\
$Y_{3,1}$  & 0.1238  & 0.1238   \\
$Y_{3,2}$  & 0.0462  & 0.0462   \\
$Y_{3,3}$  & -0.1045 & -0.1045  \\
\enddata
\tablenotetext{a}{The $Y_{\ell,m}$ are all normalized so that $Y_{0,0}$=1.00.}
\tablenotetext{b}{From \citet{addison2019minervaAustrialisFirstResults}, arXiv v1}
\tablecomments{Orbital and physical parameters of the input planets for forward models.}
\end{deluxetable}

\section{Mapping with a GCM Forward Model}\label{sec:gcmForward}

\begin{figure*}
\gridline{
	\fig{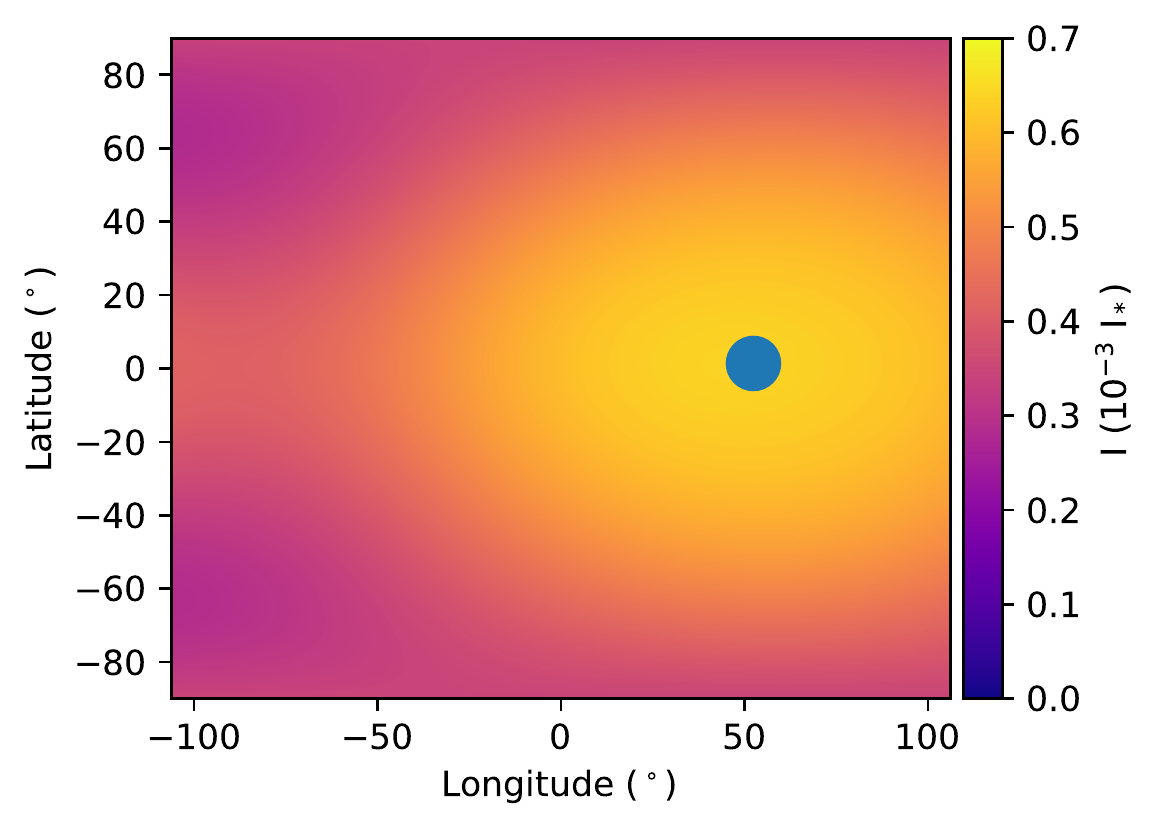}{0.3\textwidth}{Forward Input Map from GCM}
	}
\gridline{
	\fig{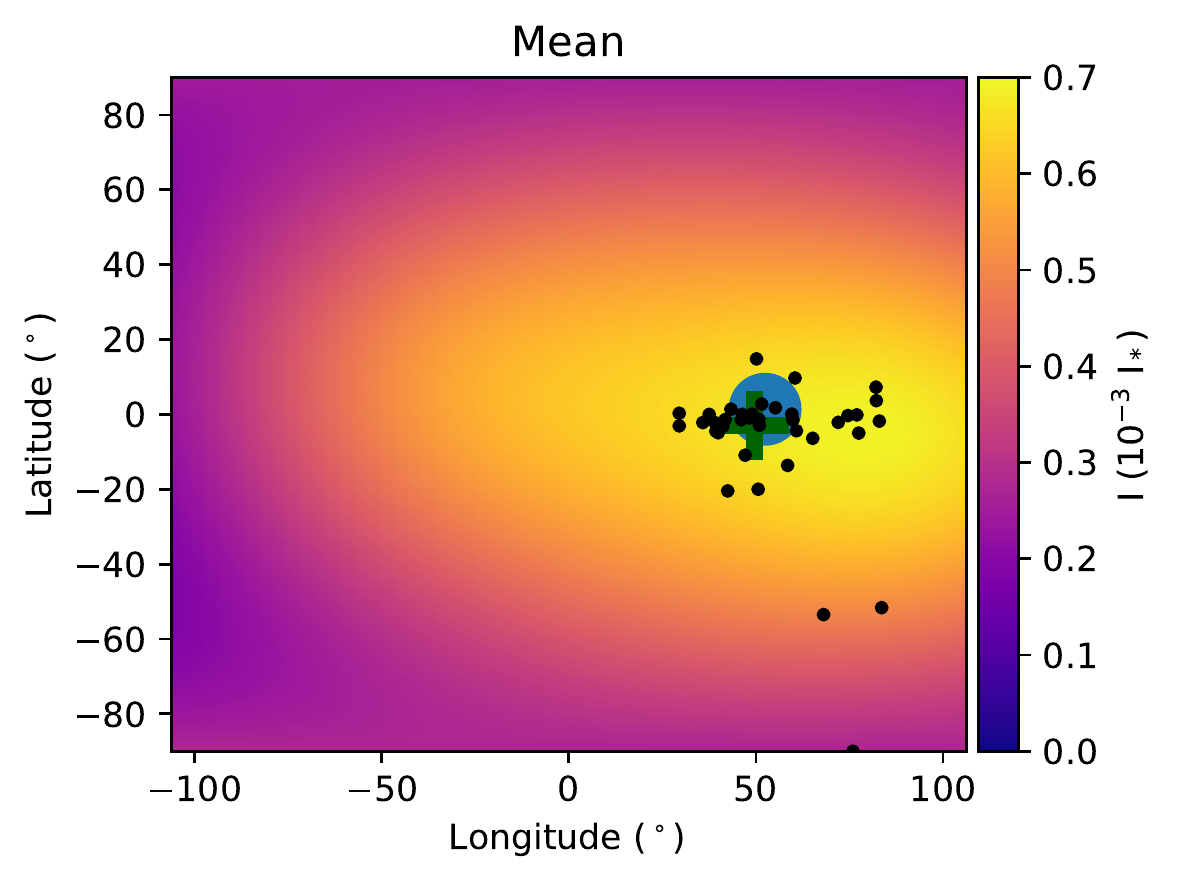}{0.30\textwidth}{Recovered \starry\ Mean with Flat Baseline}
	\fig{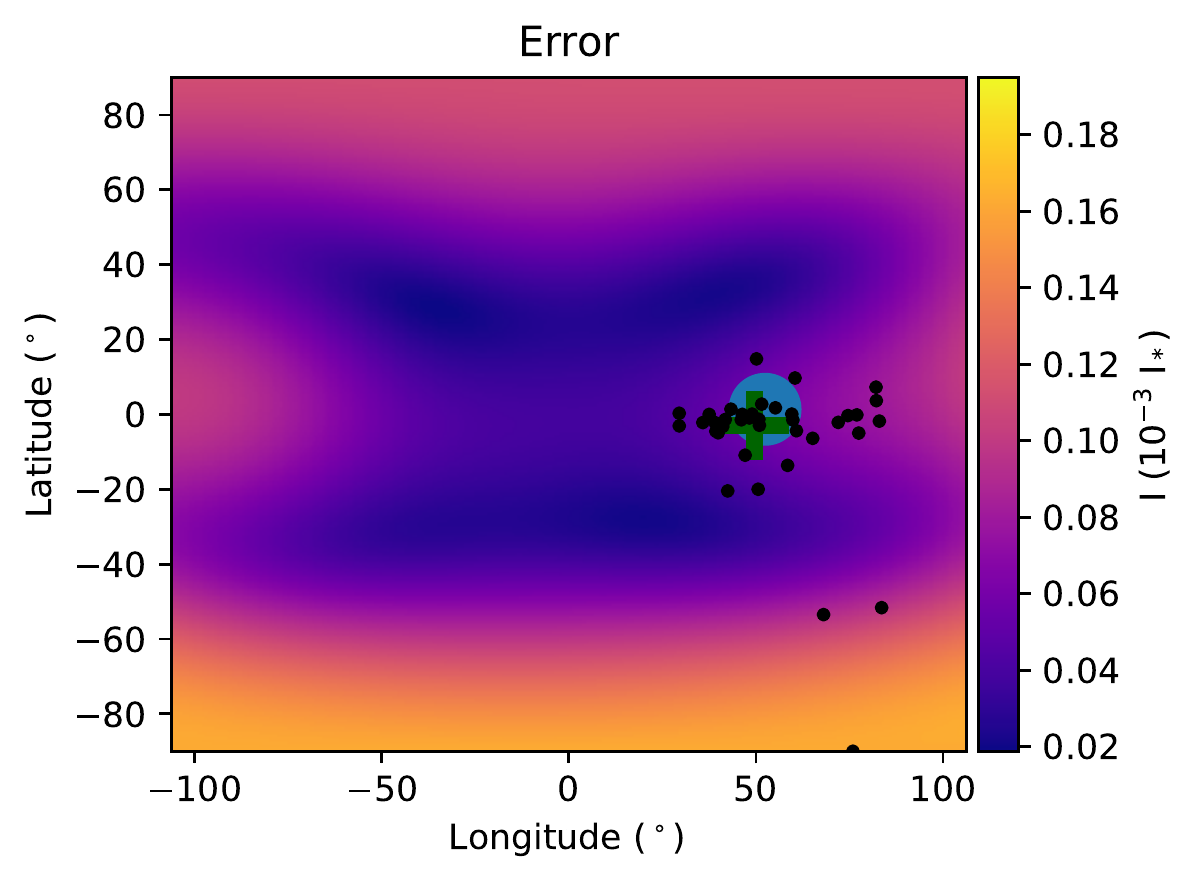}{0.30\textwidth}{Recovered \starry\ Uncertainty with Flat Baseline}
	\fig{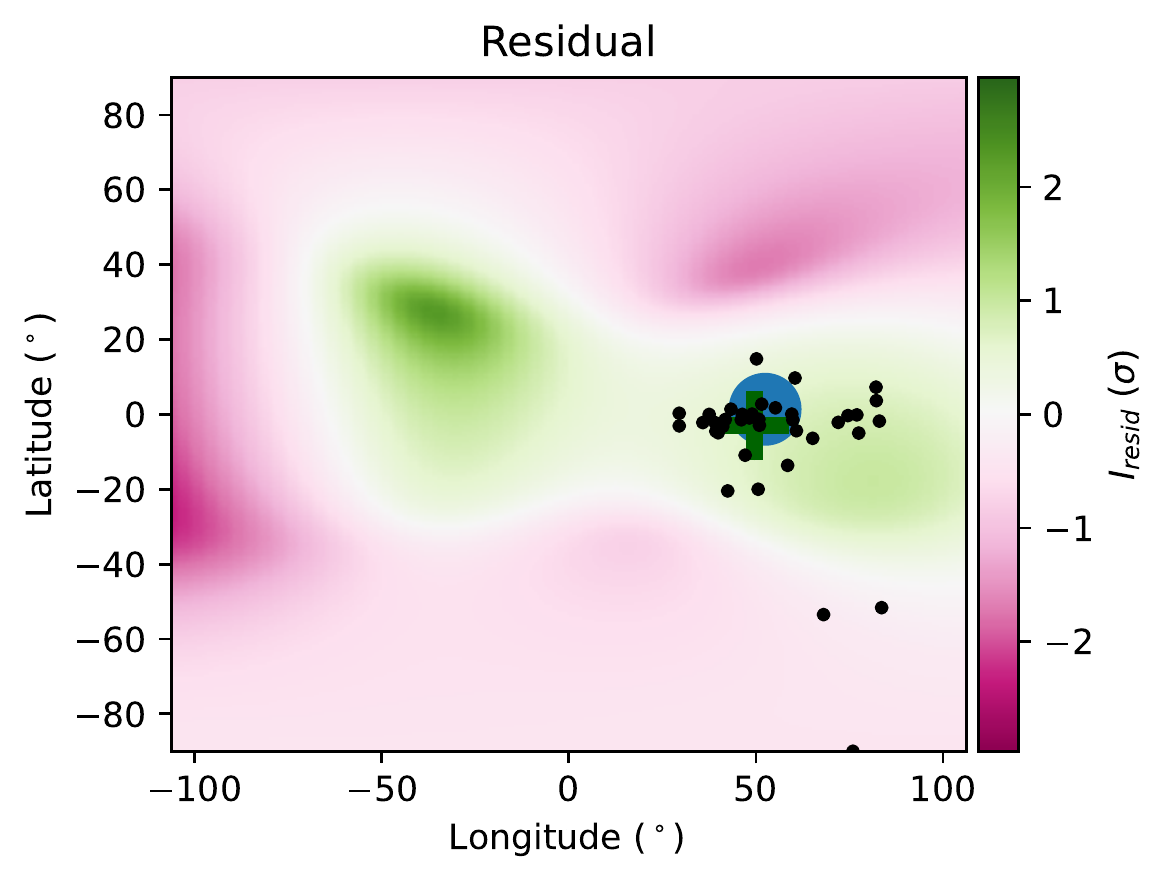}{0.30\textwidth}{Residuals with Flat Baseline}
	}
\gridline{
	\fig{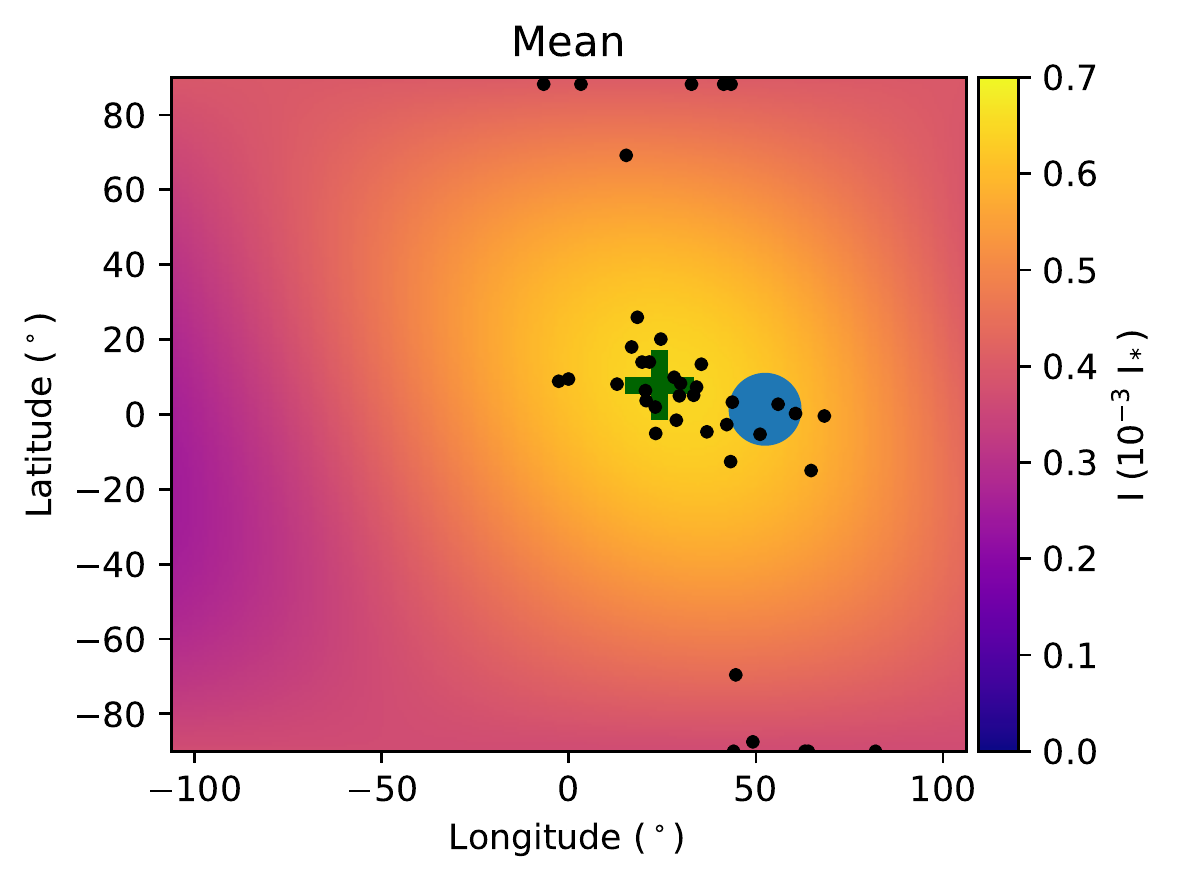}{0.30\textwidth}{Recovered \starry\ Mean with Cubic Baseline}
	\fig{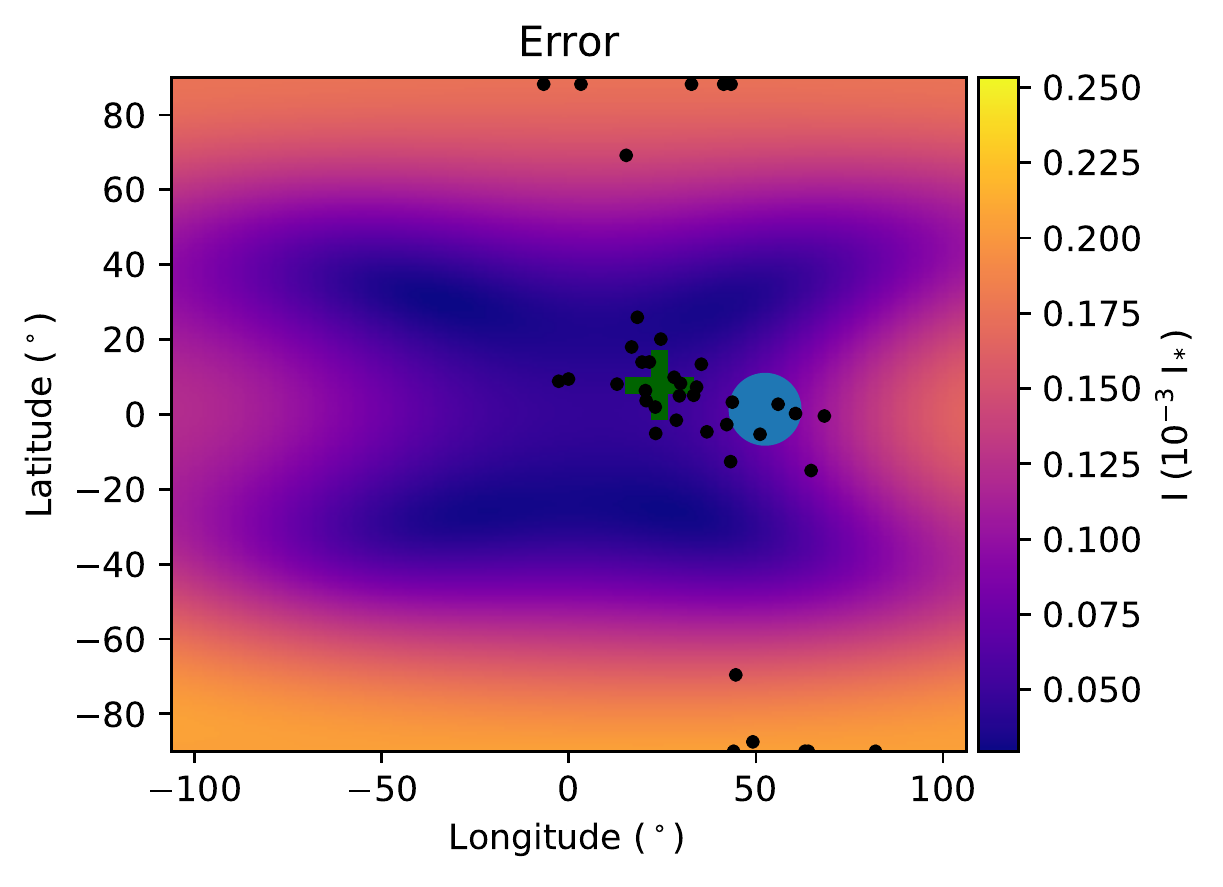}{0.30\textwidth}{Recovered \starry\ Uncertainty with Cubic Baseline}
	\fig{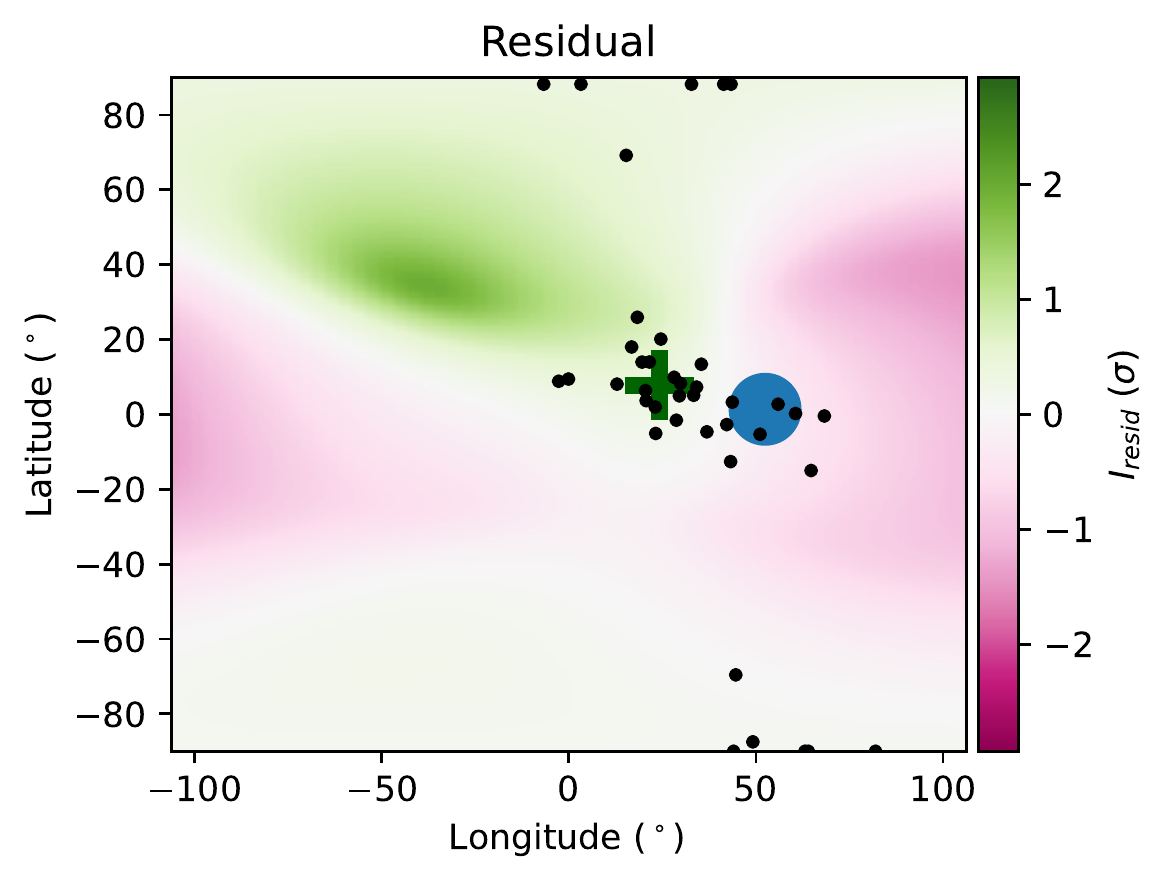}{0.30\textwidth}{Residuals with Cubic Baseline}
	}
\caption{{\it Top:} Forward map model calculated from a GCM for \hdb.
{\it Middle:} \starry\ map mean, uncertainty and residuals for \hdb\ where the baseline is flat.
{\it Bottom:} \starry\ map mean, uncertainty and residuals for \hdb\ with a cubic polynomial added to the forward model and then fit with a Gaussian process.}\label{fig:mapPosteriorsGCM}
\end{figure*}

\begin{figure*}
\gridline{	
\fig{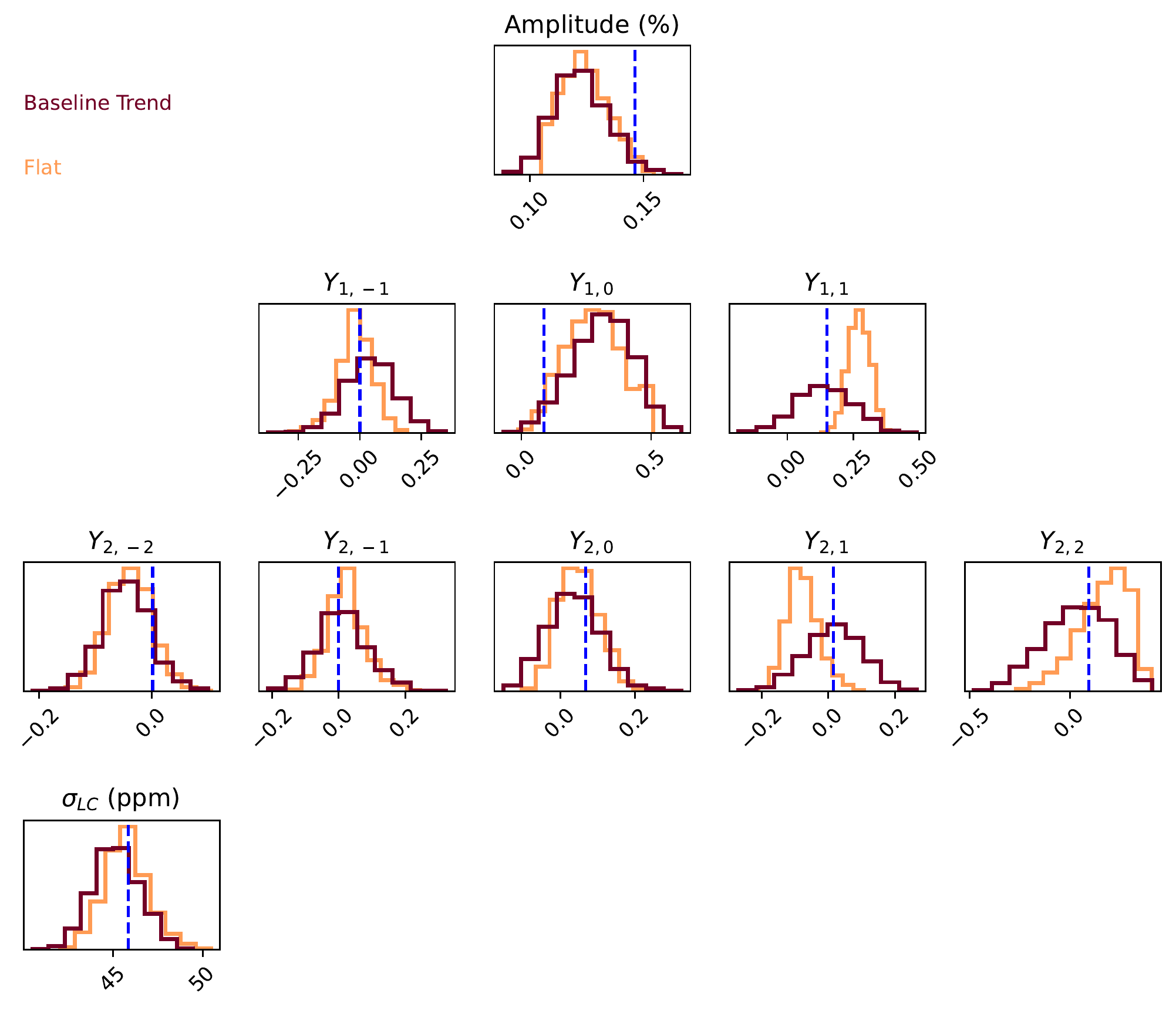}{0.99\textwidth}{Posterior Distributions for A Simulated \hdb\ Observation Using a GCM Forward Map}
	}
\caption{Posterior distributions for all mapping variables and the lightcurve standard deviation for a GCM forward map.
Results are similar to Figure \ref{fig:posteriorHistHD189}, where a different forward map was used.\label{fig:posteriorHistHD189GCM}}
\end{figure*}

We also use a General Circulation Model (Isaac Malsky, private communication) for HD 189733 b as a more realistic forward map than the arbitrary map presented in Section \ref{sec:forwardModel}.
The forward map is shown in Figure \ref{fig:mapPosteriorsGCM} fit up to spherical harmonic degree 3.
The simulated data is described in Section \ref{sec:differentForwardMap} and the resulting fits up to spherical harmonic degree 2 are shown in Figure \ref{fig:posteriorHistHD189GCM}.

\bibliographystyle{apj}
\bibliography{this_biblio}



\end{document}